\newif\iffigs\figstrue

\documentclass[paper, 12pt, letterpaper]{JHEP}

\def\Bbb{\bf}
\def\C{{\Bbb C}}

\def\Z{{\Bbb Z}}

\def\bearray{\begin{eqnarray}}
\def\eearray{\end{eqnarray}}
\def\bearraynn{\begin{eqnarray*}}
\def\eearraynn{\end{eqnarray*}}
\def\bfig{\begin{figure}}
\def\efig{\end{figure}}

\def\opeq#1{\advance\lineskip#1 \advance\baselineskip#1
        \advance\lineskiplimit#1}

\newtheorem{Proposition}{Proposition}[section]

\newtheorem{Theorem}{Theorem}[section]
\newtheorem{Lemma}{Lemma}[section]
\newtheorem{Corrolary}{Corrolary}[section]

\newcommand{\be}{\begin{equation}}
\newcommand{\ee}{\end{equation}}
\newcommand{\bea}{\begin{eqnarray}}
\newcommand{\eea}{\end{eqnarray}}

\newcommand{\bp}{\begin{Proposition}}
\newcommand{\ep}{\end{Proposition}}
\newcommand{\bt}{\begin{Theorem}}
\newcommand{\et}{\end{Theorem}}
\newcommand{\bl}{\begin{Lemma}}
\newcommand{\el}{\end{Lemma}}
\newcommand{\bc}{\begin{Corrolary}}
\newcommand{\ec}{\end{Corrolary}}
\newcommand{\nn}{\nonumber}

\usepackage{graphics}

\title{Graded Lagrangians, exotic topological D-branes and enhanced 
triangulated categories }

\author{C.~I.~Lazaroiu\\C.~N.~Yang Institute for Theoretical Physics\\
SUNY at Stony BrookNY11794-3840, U.S.A.\\calin@insti.physics.sunysb.edu}

\abstract{I point out that (BPS saturated) A-type D-branes 
in superstring compactification 
on Calabi-Yau threefolds  correspond to 
{\em graded} special Lagrangian submanifolds, a particular case 
of the graded Lagrangian submanifolds considered by M. Kontsevich and P. Seidel. 
Combining this with the categorical formulation of cubic string field theory 
in the presence of D-branes, I consider a collection of {\em topological} 
D-branes wrapped over the same Lagrangian cycle and
{\em derive} its string field action from first string-theoretic principles. 
The result is a {\em $\Z$-graded} version of super-Chern-Simons field theory 
living on the Lagrangian cycle, whose relevant string field is a degree one 
superconnection in a $\Z$-graded superbundle, 
in the sense previously considered in 
mathematical work of J.~M.~Bismutt and J.~Lott. 
This gives a refined (and modified) 
version of a proposal previously made by C. Vafa. 
I analyze the vacuum deformations 
of this theory and relate them to topological 
D-brane composite formation, upon using the 
general formalism developed in a previous paper. This allows me to identify a 
large class of topological D-brane composites (generalized, or `exotic' 
topological D-branes) which do not admit a traditional description. 
Among these are objects 
which correspond to the `covariantly constant sequences of flat bundles' 
considered by Bismut and Lott, as well as more general structures, 
which are related to the enhanced triangulated 
categories of Bondal and Kapranov.
I also give a rough sketch of the relation  
between this construction and the large radius limit of a certain 
version of the `derived category of Fukaya's category'. 

This paper forms part of a joint project with Prof. 
S.~Popescu, a brief announcement of which can be found in the second 
part of the note hep-th/0102183. 
The paralel B-model realization, as well as the relation with 
the enhanced triangulated categories of Bondal and Kapranov, was 
recently discussed by D.~E.~Diaconescu in the paper hep-th/0104200, upon 
using the observations contained in that announcement.
}

\begin{document}

\tableofcontents

\pagebreak

\section{Introduction}

Recent progress in our understanding of Calabi-Yau D-brane physics 
has lead to the realization that the D-branes of an $N=2$ superstring 
compactification are intrinsically graded objects. This observation, 
which in the physics context is due to M. Douglas \cite{Douglas_Kontsevich,
pi_stab}, and in the homological mirror symmetry literature can be traced 
back to \cite{Kontsevich} (see also \cite{Seidel} and \cite{Zaslow_hms}), 
has far reaching 
implications for the physical description of open superstring dynamics 
on Calabi-Yau backgrounds. 
Most work toward extracting the physical consequences of this 
fact \cite{Douglas_Kontsevich, Aspinwall, Diaconescu, Oz_triples},  
has focused on the analysis of so-called B-type branes \cite{Ooguri}, 
i.e. those D-branes which in the large radius limit are described by 
holomorphic sheaves. The other class of D-branes which can be 
introduced in such compactifications (the so-called A-branes, 
which correspond to (special) Lagrangian 
cycles) has been comparatively less well studied.
This asks for clarification, 
especially since applications to mirror symmetry 
(which interchanges the two types of branes) require 
that we understand both sides of the mirror duality. 

Since the chiral/antichiral 
primary sectors of the compactified superstring are 
faithfully described by the associated topological models
\cite{Witten_NLSM, Witten_mirror, Witten_CS}, it is 
natural to approach this problem from the point of view of topological 
string theory. The purpose of 
the present paper is to analyze some basic 
dynamical implications of the existence of D-brane grading 
for the topological branes of the A model.

Most recent work on categories of D-branes on Calabi-Yau manifolds 
follows the largely {\em on-shell} (or rather, partially off-shell) 
approach originally proposed in 
\cite{Douglas_Kontsevich}. In \cite{com3}, I proposed an alternative, 
{\em consistently off-shell} approach, which makes direct use of a formalism
(developed in \cite{com1}) 
for associative string field theory in the presence of D-branes
and goes well-beyond the boundary 
state formalism \cite{Cardy, boundary_states} 
which has traditionally dominated the subject\footnote{Some limitations of the 
(topological) boundary state formalism were discussed in \cite{top}.}.
This method is especially well-suited for analysis of D-brane
composite formation, a phenomenon which forms the crucial ingredient 
of any realistic approach to Calabi-Yau D-brane physics. Since condensation 
of boundary and boundary condition changing operators is intrinsically 
an {\em off-shell} process, it can be expected  
that a complete understanding of 
such phenomena can only be obtained by systematic use of 
off-shell techniques, which 
explains the relevance of string field theory methods.
This point of view also allows for better contact with the 
homological mirror symmetry conjecture of \cite{Kontsevich}, which is based 
on an effort to work at the (co)chain, rather than (co)homology level.

In this paper I shall follow the approach outlined in \cite{com3} 
by giving its concrete realization for a class of topological 
D-branes of the A model. A thorough understanding of A-model open string dynamics 
requires consideration of intersecting D-branes and a detailed analysis of 
disk instanton effects, which ultimately should be carried out through 
mirror symmetry. Since a  complete discussion of  these issues (which bear 
an intimate relation 
with the mathematical work of K. Fukaya \cite{Fukaya, Fukaya2}) 
is rather involved, the present paper restricts to a small sector of the 
relevant structure, by considering a system of D-branes which wrap 
a given (special) Lagrangian cycle. We shall moreover take
the large radius limit, thereby neglecting disk 
instanton corrections \cite{boundary} to the classical string field action. 
As we shall see below, even the 
analysis of this sector is considerably more subtle than has previously 
been thought. In fact, insistence on a consistent off-shell 
description leads to novel results, which 
would be difficult to extract through the `mixed' methods of 
\cite{Douglas_Kontsevich, Aspinwall}. Among these is is the fact 
that the string field associated with such boundary sectors 
is a superconnection living in a 
{\em $\Z$-graded} superbundle, rather that the standard $\Z_2$-graded variant 
of \cite{Quillen} which was used in \cite{Oz_superconn, Vafa_cs}.  
More importantly,  we shall {\em show the existence of 
many more classes of {\em topological} 
A-type branes than has previously been suspected}. 
Such exotic topological branes correspond to backgrounds 
in the {\em extended} moduli space of the open topological A-string.
Whether they also play a role in the physical, untwisted theory is 
a question which we do not attempt to settle in this paper.

The paper is organized as follows. In Section 2, we recall the basics of 
the general 
formalism of associative open string field theory with D-branes which 
was developed in \cite{com1, com3}. Since we shall later deal with the 
A-model, which for certain D-brane configurations admits a 
`complex conjugation' symmetry, we also give a brief discussion of the 
supplementary structure describing open string field 
theories endowed with such operations. This is a straightforward 
extension of the analysis of \cite{com1}, which bears on basic structural 
issues such as consistent general constructions of antibranes. It can be 
viewed as a D-brane extension of the analysis already given in 
\cite{Witten_SFT, Thorn, Gaberdiel}, 
though our discussion is carried in different conventions. 
In Section 3 I reconsider the problem of grading of A-type topological 
D-branes. While this issue was touched upon in \cite{Douglas_Kontsevich} 
(see also \cite{Aspinwall}), a careful geometric analysis does not seem to 
have been given before. I give a precise description of this grading 
and discuss some of the underlying issues of orientation, which turn out 
to be of crucial importance for the string field theory discussed in later 
sections. Our approach makes us of the 
so-called {\em graded Lagrangian submanifolds} of \cite{Seidel},  which 
can be shown to give the general description of 
(not necessarily BPS saturated) graded topological A-branes. Since the 
present paper 
restricts to the special Lagrangian case, I only discuss this theory 
in its very simplified form which applies to such situations. 
This gives a geometrically self-contained description of A-type branes 
as graded objects, thereby improving on the discussion 
of \cite{Douglas_Kontsevich, Aspinwall}. The general theory, as it applies 
to non-BPS A-type branes, will be discussed somewhere else \cite{nlsm}. 
Armed with a consistent off-shell framework and a 
precise understanding of the geometric description of our 
objects, we proceed in Section 4 to construct the string field theory 
of a system of {\em distinct} D-branes wrapping the same special Lagrangian 
cycle. More precisely, we consider a set of branes having 
different gradings and whose underlying geometric supports coincide. 
The fact that one must consider such systems is a direct consequence of 
the idea of D-branes as graded objects, and it should serve as a  
test for its implications. The salient point of our construction is 
that the presence of distinct gradings leads to a shift of the worldsheet 
$U(1)$ charge of the various boundary condition changing observables, 
which implies that the resulting string field theory is a {\em $\Z$-graded} 
version of super-Chern-Simons theory
\footnote{This should be compared with 
the $\Z_2$-graded proposal of \cite{Vafa_cs}.}. 
This suggests that the currently prevalent approach
(which is largely based on borrowing the results of 
\cite{Witten_CS}) must be reconsidered. 
We build the string field theory by identifying each piece of the 
axiomatic data discussed in \cite{com1, com3}, and check the relevant 
consistency constraints. This gives an explicit 
realization  of the general framework developed in those papers. 
After identifying the underlying structure, we show that the resulting 
string field action admits a description in the language of 
{\em $\Z$-graded superconnections}, which were previously considered in the
mathematical work of \cite{Bismut_Lott}. 
We proceed in 
Section 5 with a discussion of the conditions under 
which our string field theory 
is invariant with respect to complex conjugation, and give a precise 
description
of the conjugation operators. 
Section 6 formulates an extended string field action, whose detailed 
analysis is left for future work. Section 7 gives
a preliminary analysis of the moduli space of vacua of our string field 
theory. Upon applying the general framework of \cite{com1, com3}, we 
discuss the relevant deformation problem and sketch its relation with 
the modern mathematical theory of {\em extended} deformations 
\cite{Kontsevich_Felder, Manin}. This provides a concrete realization
of some general observations made in \cite{com3}. 
In Section 8 we analyze various 
types of deformations, which allows us to 
identify large classes of topological A-type branes 
(wrapping the cycle $L$) which {\em do not admit a traditional description}. 
These `exotic A-type branes' can be viewed as topological D-brane composites 
resulting from condensation of boundary and boundary condition changing 
operators, and they must be included in a physically complete analysis of 
A-type open string dynamics. As a particular case, we recover the standard 
deformations of traditional A-type branes, and a class of generalized 
D-branes which correspond to the flat complexes of vector bundles 
studied in \cite{Bismut_Lott}. We also discuss more 
general solutions, which correspond to the pseudocomplexes and generalized 
complexes of \cite{com1}, and relate the former to the enhanced triangulated 
categories of Bondal and Kapranov, upon following the general observations 
already made in \cite{com3}. 
This provides a vast enlargement of the theory of 
\cite{Witten_CS}, which can be analyzed through 
the categorical methods developed in 
\cite{com1}.  Section 9 makes some brief remarks 
on how the theory 
considered in this paper relates to Fukaya's 
category \cite{Fukaya, Fukaya2}. Finally, section 10 
presents our conclusions.

The content of this paper was originally conceived as 
introductory material for the more detailed work \cite{us}, 
a brief announcement of 
which can be found in the last section of the note \cite{com3}. 
Meanwhile, a paper appeared \cite{Diaconescu}, which succeeded to recover some 
of the B-model details which were left out in \cite{com3}, as well as 
the specifics of the relation with the work of Bondal and Kapranov \cite{BK}, 
a relation which was mentioned in \cite{com3}. 
This prompted me to write the present note, which 
contains a self-contained description of (part) of the A-model realization 
of the relevant string field theory, and a brief 
sketch of the ensuing mathematical 
analysis. Since the basic  B-model realization is
now available in \cite{Diaconescu}, I refer the reader to 
this reference for the parallel holomorphic discussion
\footnote{I should perhaps point out that the idea of using the work 
of Bondal and Kapranov \cite{BK} goes back to the original paper 
of M. Kontsevich on the homological mirror symmetry conjecture. This 
idea was re-iterated in \cite{com3} in a string field theory context, 
where its application to the B model was mentioned without giving 
all of the relevant 
details (which form part of the more ambitious project \cite{us}). The paper 
\cite{Diaconescu} follows this idea by studying a $\Z$-graded 
holomorphic theory, which was implicit (but not explicitly written down) 
in the announcement \cite{com3}. That theory, as well as the theory of this 
paper, can be obtained through the very general procedure 
of {\em shift-completion}.}.

\section{Review of axioms and complex conjugation}

We start with a short review of the necessary framework, 
followed by a brief discussion of complex conjugations. 
We are interested in a general description of the structure of 
cubic (or associative, as opposed to homotopy associative) 
open string field theory in the presence of D-branes. This subject
was discussed systematically in the paper \cite{com1}, to which we refer 
the reader for details. The basic idea is to formulate the theory in 
terms of a so-called `differential graded (dG) category', which encodes 
the spaces of off-shell states of strings stretching between 
a collection of D-branes. Since we are interested in oriented strings, 
such states can be viewed as morphisms between the various D-branes, 
which allows us to build a category with objects given by the 
D-branes themselves. The associative morphism compositions are given by the 
basic string products, which result from the triple correlators 
on the disk. From this perspective, states of strings whose endpoints lie on the 
same D-brane $a$ define `diagonal' boundary sectors $Hom(a,a)=End(a)$, while 
states of strings stretching between two different D-branes $a$ and $b$ 
(namely from $a$ to $b$) define boundary condition changing sectors $Hom(a,b)$. 
This terminology is inspired by the formalism of open-closed conformal 
field theory on surfaces with boundary \cite{Cardy}; indeed, 
states in the boundary  and 
boundary condition changing sectors are related to boundary/boundary condition changing 
conformal field theory operators via the state-operator correspondence.
The reader is referred to \cite{com3} for a nontechnical introduction 
to this approach and to \cite{com1} for a more detailed discussion. 
Some background on dG categories can be found in the appendix of \cite{com1}, 
while a systematic discussion of the relevant on-shell approach can be 
found in \cite{top}.

\subsection{The basic data and the string field action} 

Recall from \cite{com1} that a 
(tree level) associative open string field theory in the presence of D-branes 
is specified by: 

\

(I) A differential graded $\C$-linear category ${\cal A}$

\

(II) For each pair of objects $a, b$ of ${\cal A}$, an invariant 
nondegenerate bilinear and graded-symmetric form 
$_{ab}(.,.)_{ba}:Hom(a,b)\times Hom(b,a)\rightarrow \C$ of 
degree $3$.

\

A {\em $\C$-linear category} is a category whose morphism spaces are 
complex vector spaces and whose morphism compositions are bilinear maps. 
A {\em graded linear category} is a linear category whose morphism spaces 
are $\Z$-graded, i.e. $Hom(a,b)=\oplus_{k\in \Z}{Hom^k(a,b)}$, and 
such that morphism compositions are homogeneous of degree zero, i.e.: 
\be
|uv|=|u|+|v|~~{\rm~for~all~homogeneous~}u\in Hom(b,c),~v\in Hom(a,b)~~,
\ee 
where $|.|$ denotes the degree of a homogeneous element.
In a {\em differential graded linear category} (dG category), 
the morphism spaces 
$Hom(a,b)$ are endowed with nilpotent operators $Q_{ab}$  
of degree $+1$ which act as derivations of morphism compositions:
\be
\label{der}
Q_{ac}(uv)=Q_{bc}(u)v+(-1)^{|u|}uQ_{ab}{v}~{\rm~for~homogeneous~}u\in Hom(b,c),~
v\in Hom(a,b)~~.
\ee
Graded symmetry of the bilinear forms means: 
\be
_{ab}(u,v)_{ba}=(-1)^{|v||u|}_{ba}(v,u)_{ab}
\ee 
for homogeneous elements $u\in Hom(a,b)$ and $v\in Hom(b,a)$. 
The degree $3$ constraint is:
\be
_{ab}(u,v)_{ba}=0~~{\rm~unless~}|u|+|v|=3~~.
\ee
The bilinear forms are required to be invariant with respect 
to the action of $Q_{ab}$ and morphism compositions:
\be
\label{d_invariance}
_{ab}(Q_{ab}(u),v)_{ba}+(-1)^{|u|}(u,Q_{ba}(v))=0~~{\rm~for~}
u\in Hom(a,b),~v\in Hom(b,a)~~
\ee
and
\be
_{ca}(u,vw)_{ac}=_{ba}(uv,w)_{ab}~~{\rm~for~}u\in Hom(c,a),~v\in Hom(b,c),~
w\in Hom(a,b)~~.
\ee

Given such data, one can build the (unextended) string field action:
\be
\label{gfeneral_action_expanded}
S(\phi)=\frac{1}{2}\sum_{a,b\in S}
{_{ba}\langle \phi_{ba} , Q_{ab}\phi_{ab} \rangle_{ab}} +
\frac{1}{3}\sum_{a,b,c\in S}{
_{ca}\langle \phi_{ca}, \phi_{bc}\cdot \phi_{ab}\rangle_{ac}}~~, 
\ee
where $S$ denotes the set of objects of ${\cal A}$\footnote{We assume for 
simplicity that ${\cal A}$ is a small category, i.e. its objects form a set.
In fact, we shall only need the case when $S$ is finite or countable, since this 
will be relevant for our application.
We neglect the conditions under which various sums 
make mathematical sense (i.e. converge etc). Convergence of 
these sums can be assured (at least in principle) 
by imposing supplementary conditions on the 
allowed string field configurations.}
and $\phi_{ab}\in Hom^1(a,b)$ are the components of the degree one string 
field.

Upon defining the total boundary space ${\cal H}=\oplus_{a,b\in S}{Hom(a,b)}$, 
the total string product $\cdot:{\cal H}\times {\cal H}\rightarrow {\cal H}$
as well as the total bilinear form $\langle .,.\rangle$ 
and BRST operator $Q:{\cal H}\rightarrow {\cal H}$ in an obvious manner 
(see \cite{com1} for details), one can write this action in the more compact 
form:
\be
\label{general_action}
S(\phi)=\frac{1}{2}\langle \phi , Q\phi \rangle +\frac{1}{3}
\langle \phi, \phi\cdot \phi\rangle~~~,
\ee
where $\phi=\oplus_{a,b\in S}{\phi_{ab}}$ is a degree one element of 
${\cal H}$. 

\subsection{Theories with complex conjugation} 

The axioms of a string field theory can be supplemented 
by requiring the existence of conjugation operators 
subject to certain constraints\footnote{This type of structure was 
discussed in \cite{Gaberdiel} for the case of a single boundary sector
(but allowing for homotopy associative string products).}. 
We say that such a theory is {\em endowed with conjugations}
if one is given the following two supplementary structures:

(IIIa) An involutive map $a\rightarrow {\overline a}$ on the set of 
D-brane labels.

(IIIb) A system of {\em degree zero} antilinear operators 
$*_{ab}:Hom(a,b)\rightarrow Hom({\overline b},{\overline a})$, 
with the properties:

\

(1) $*_{{\overline a}{\overline b}}*_{ba}=
id_{Hom(b,a)}$, for any two objects $a$ and $b$

\

(2) $_{{\overline b}{\overline a}}\langle *_{ab}u,*_{ba}v
\rangle_{{\overline a}{\overline b}}=
_{ba}\overline{\langle v, u\rangle}_{ab}$ for $u\in Hom(a,b)$ and 
$v\in Hom(b,a)$

\

(3) $Q_{{\overline b}{\overline a}}*_{ab}u=(-1)^{|u|+1}*_{ab}Q_{ab}u$ for 
$u\in Hom(a,b)$

\

(4) $*_{ac}(u_{bc}v_{ab})=(*_{ab}v_{ab})(*_{bc}u_{bc})$ for 
$u_{bc}\in Hom(b,c)$ and $v_{ab}\in Hom(a,b)$. 

\

For each D-brane $a$, its partner ${\overline a}$ will be called its 
`conjugate brane'.

In a theory possessing conjugations, the action (\ref{general_action}) has the 
following property:
\be
\overline{S(\phi)}=S(*\phi)~~.
\ee
Hence one can assure reality of 
the string field action  by imposing the following condition on the 
string field:
\be
*\phi=\phi\Leftrightarrow *_{ab}\phi_{ab}=
\phi_{{\overline b}{\overline a}}~~.
\ee
The condition that $*$ preserve the degree $|.|$ is crucial for 
consistency of the reality constraint $|*\phi|=|\phi|$ with the 
degree one constraint $|\phi|=1$.

\section{A-type D-branes as graded special Lagrangian submanifolds}

It is well-known \cite{Ooguri} that BPS saturated D-branes in Calabi-Yau threefold 
compactifications are described  either by holomorphic cycles 
(so-called type B branes) or by 
special Lagrangian cycles (so-called type A D-branes) 
of the Calabi-Yau target space $X$. 
In the case of multiply-wrapped branes one must also include a bundle 
living on each cycle (which corresponds to a choice of Chan-Paton data) 
and a choice of connection in this bundle, which should be 
integrable for B-type branes and flat for type A branes. 
From a conformal field theory point of view, type A and B D-branes 
correspond to different boundary conditions \cite{Ooguri}, and they 
preserve 
different $N=2$ subalgebras of the $(2,2)$ worldsheet algebra.

The starting point of our analysis is the observation of 
\cite{Douglas_Kontsevich} 
that this data does not in fact suffice for a complete description of D-brane 
physics. 
The essence of the argument of \cite{Douglas_Kontsevich} 
is as follows. The various boundary/boundary 
condition changing sectors of the theory consist of open string states, which 
are charged with respect to the worldsheet $U(1)$ current of the $N=2$ 
superconformal algebra. It is then shown in \cite{Douglas_Kontsevich} 
(based on bosonization techniques) 
that a complete specification of the theory requires 
a consistent choice of charges for the various boundary 
sectors, and that, 
in the presence of at least two different D-branes, the 
relative assignment of such charges has 
an invariant physical meaning. If one defines the `abstract' degree 
of a D-brane through the winding number of the bosonized $U(1)$ current, 
this amounts to the statement that a complete 
description of the background requires the specification of an 
integer number for each D-brane present in the compactification.

While this is an extremely general argument (which in particular 
applies to non-geometric compactifications), the concrete realization 
for the A model remains somewhat obscure. In this section, I 
explain the geometric meaning of this `grade' for the case of semiclassical 
type A branes, i.e. for the large radius limit of a Calabi-Yau 
compactification which includes such objects\footnote{If one wishes to go away from 
large radius, then one must consider disk 
instanton corrections to the boundary/boundary condition changing sectors, 
which are induced by corrections to the associated BRST operators. 
This leads to Floer cohomology and instanton destabilization of some semiclassical 
D-branes, as well as to supplementary corrections to the string field action. 
By staying in the large radius limit, we avoid each of these issues.}.
Our proposal is motivated by a combination of mathematical results of 
\cite{Seidel} and an anomaly analysis which can be carried out in the twisted 
model (the A-model). Since a complete exposition would take more space 
than afforded in this paper, I will give a simplified discussion and refer the 
reader to \cite{nlsm}, which will include a more detailed analysis. 

We propose that the correct description of a BPS saturated 
type A brane is given 
(in the large radius limit) by a triple $({\bf L}, E, A)$, where 
${\bf L}$ is a so-called {\em graded special Lagrangian submanifold} of $X$.
The mathematical 
concept of graded Lagrangian manifolds (not necessarily special) 
is originally due to M. Kontsevich 
\cite{Kontsevich} and was discussed in more detail in recent work of 
P. Seidel \cite{Seidel}. Recall that a Lagrangian cycle is a three-dimensional 
submanifold of $X$such that the Kahler form $\omega$ of $X$ 
has vanishing restriction to $L$. 
The cycle is {\em special} Lagrangian, if also also has:
\be
\label{sl}
Im(\lambda_L\Omega)|_L=0~~
\ee
for some complex number $\lambda_L$ of unit modulus.
Since this condition 
is invariant under the change $\lambda_L\rightarrow -\lambda_L$, the relevant 
quantity is $\lambda_L^2$, i.e. the special Lagrangian condition (\ref{sl}) 
only specifies $\lambda_L$ {\em up to sign}.
A {\em grading}\footnote{It is 
not hard to show \cite{nlsm} that this is a particular case of the general 
notion of graded Lagrangian submanifold introduced in \cite{Seidel}.}  
of a special Lagrangian cycle $L$ 
is simply the choice of a {\em real} number $\phi_L$ such that 
$e^{-2\pi i \phi_L}=\lambda_L^{2}$.
There is a discrete infinity of such 
choices (differing by an integer), so there is a countable infinity 
of gradings of any given special Lagrangian cycle. 
Furthermore, there is 
always a {\em canonical} choice $\phi_L^{(0)}\in [0,1)$ (which we shall 
call the {\em fundamental grading}), a fact which 
distinguishes special Lagrangians from more general Lagrangian submanifolds. 
Then any grading is of the form $\phi_L^{(n)}=\phi_L^{(0)}+n$, with $n$ an
integer which specifies a shift of the charges of the associated 
boundary  sector. Hence one can identify a graded special Lagrangian 
cycle ${\bf L}$ with the pair $(L,n)$. This agrees with the general conformal 
field theory argument of \cite{Douglas_Kontsevich}.

We end this section by noting that a choice of grading specifies an 
orientation of the special Lagrangian cycle. Indeed, given a grading 
$\phi_L^{(n)}$, one has a canonically-defined 
{\em real} volume form $\Omega_L(n):=e^{-i\pi \phi_L^{(n)}}\Omega=
e^{-i(\phi_L^{(0)}+n)}\Omega$ 
on $L$\footnote{One can also think of this as a choice of sign, 
$\lambda_L(n):=e^{-i\pi \phi_L(n)}$, for the quantity appearing in (\ref{sl})
(the other choice afforded by the relation $\lambda_L^2=
e^{-2\pi i \phi_L^{(n)}}$
would be $\lambda'_L(n)=-\lambda_L=e^{-i\pi \phi_L^{(n+1)}}$, which 
corresponds to the opposite orientation). The point is that once one 
specifies a grading $n$, we have a choice $\lambda_L(n)$ which is 
uniquely determined by $n$, and consequently we have a 
natural choice of orientation. In the absence of a grading, we would have 
no natural way to pick one of the two opposite orientations. This may seem 
subtle but it is in fact a triviality. As we shall see below, this 
simple fact is ultimately responsible for the appearance of supertraces in our 
string field action. }.
It is clear that the orientation induced by $\phi_L^{(n)}=\phi_L^{(0)}+n$ 
in this manner coincides with 
$(-1)^n$ times the orientation induced by the fundamental grading 
$\phi_L^{(0)}$.

\section{The open string field theory of a collection of 
graded D-branes wrapping the same 
special Lagrangian cycle}

We consider a family of D-branes $a_n$ described by triples  
$({\bf L}_n, E_n, A_n)$ 
with ${\bf L}_n:=(L, n)$ for some collection of integers 
$n$. These branes share the same underlying 
special Lagrangian cycle $L$, but have possibly different Chan-Patton bundles 
$E_n$ (complex vector bundles defined over the cycle $L$) and  different 
background flat connections $A_n$ living in these bundles. Note that we 
assume that each brane has a different grade $n$.
  
According to our previous discussion, each D-brane $a_n$ 
determines an 
orientation ${\cal O}_n$ of the cycle $L$, and these orientations are 
related 
to the `fundamental' orientation ${\cal O}_0$ through:
\be
{\cal O}_n=(-1)^n{\cal O}_0~~.
\ee\noindent When integrating differential forms (see below), we shall 
write $L_n$ for $L$ endowed with the orientation ${\cal O}_n$, and $L:=L_0$ 
for $L$ endowed with the orientation ${\cal O}_0$.

We now consider the cubic open string field theory associated with 
such a system. Since we have more than one D-brane, one must use the 
categorical framework of \cite{com1, com3}, which was shortly reviewed in 
Section 2. Instead of giving a lengthy discussion of localization 
starting from the topological A model, we shall simply list the relevant data
and check that the axioms of Section 2 are satisfied. The data of 
interest is as follows:

(1) The spaces $Hom(a_m,a_n)$ of off-shell 
states of oriented open strings stretching from $a_m$ to $a_n$. 
For our topological field theory, these 
can be identified through a slight extension of the arguments of 
\cite{Witten_CS}, which gives:
\be
\label{spaces}
Hom^k(a_m,a_n)=\Omega^{k+m-n}(L, Hom(E_m, E_n))~~,
\ee
where
$\Omega^*(L, Hom(E_m, E_n))$ 
denotes the space of (smooth) differential forms on the cycle 
$L$ with values in the bundle $Hom(E_m, E_n)$. In these expressions we let $k$ be any 
signed integer, so for example $Hom^k(a_m,a_n)$ can be nonvanishing for $k=n-m...n-m+3$ 
and vanishes otherwise. 

In the string field theory of the A model, the various state spaces must 
graded 
by the charge of boundary/ boundary condition changing states with respect 
to the anomalous $U(1)$ current on the string worldsheet (this 
follows from the construction of the string field products in the manner of 
\cite{Zwiebach_closed, Zwiebach_open}). 
In our situation, this grading (which is indicated by the superscript 
$k$ in (\ref{spaces})) differs from the
grading by form rank, as displayed 
by the shift through $m-n$ in (\ref{spaces}). The presence of such a 
shift follows from arguments similar 
to those of \cite{Douglas_Kontsevich}, or by a careful discussion of 
localization along the lines of \cite{Witten_CS}. 
These shifts of the $U(1)$ 
charge in the boundary condition changing sectors reflect the
different gradings of the branes $a_n$ and are required for a consistent 
description of the theory. They will play a crucial role in the correct 
identification of the string field theory of our D-brane system. 
If one denotes the $U(1)$ charge of a state $u\in Hom(a_m, a_n)$ by $|u|$, 
then one has the relation:
\be
\label{U1}
|u|=rank u+n-m~~.
\ee
Relation (\ref{spaces}) can also be written as:
\be
Hom(a_m,a_n)=\Omega(L, Hom(E_m, E_n))[n-m]~~,
\ee
upon using standard mathematical notation for shifting degrees.

It is useful to consider the total boundary state space
${\cal H}=\oplus_{m,n}{Hom(a_m, a_n)}=\Omega^*(L, End({\bf E}))$, 
where ${\bf E}=\oplus_{n}{E_n}$. This has an obvious $\Z\times \Z$
grading given by ${\cal H}_{m,n}=Hom(a_m, a_n)$, and a
$\Z_2$ grading given by:
\bea
\label{eo}
{\cal H}_{even}&=&\oplus_{k+m-n=even}{Hom^k(a_m, a_n)}~~,\\
{\cal H}_{odd}&=&\oplus_{k+m-n=odd}{Hom^k(a_m,a_n)}~~.\nn
\eea
The $\Z_2$ grading can be described in standard supergeometry language 
as follows. Let us define even and odd subbundles of ${\bf E}$ through:
\bea
\label{E_grading}
E_{even}&=&\oplus_{n=even}{E_n}~~\\
E_{odd}&=&\oplus_{n=odd}{E_n}~~.\nn
\eea
Then ${\bf E}=E_{even}\oplus E_{odd}$ can be regarded as a super-vector 
bundle ($\Z_2$-graded bundle). Its bundle of endomorphisms then has 
the decomposition:
\be
\label{end_grading}
End({\bf E})=Hom(E_{even}, E_{even})\oplus Hom(E_{odd}, E_{odd})\oplus
Hom(E_{even}, E_{odd})\oplus Hom(E_{odd}, E_{even})~~
\ee
and a total $\Z_2$ grading:
\bea
\label{superbundle}
End({\bf E})_{even}&=&
Hom(E_{even}, E_{even})\oplus Hom(E_{odd}, E_{odd})~~\\
End({\bf E})_{odd}&=&Hom(E_{even}, E_{odd})\oplus Hom(E_{odd}, E_{even})~~.\nn
\eea
Then ${\cal H}$ corresponds to the the tensor product
$\Omega^*(L)\otimes_{\Omega^0(L)} \Omega^0(End({\bf E}))$, 
in which case the $\Z_2$ grading on ${\cal H}$ is induced by the standard 
$\Z_2$ grading on $\Omega^*(L)$ and the $\Z_2$-grading (\ref{superbundle}) 
on $End({\bf E})$. 

Also note that $End({\bf E})$ has a diagonal $\Z$-grading, whose 
mod $2$ reduction gives the grading (\ref{end_grading}):
\be
\label{E_Zgrading}
End({\bf E})_{s}=\oplus_{m-n=s}{Hom(E_m,E_n)}~~.
\ee
As a consequence, the total boundary state space ${\cal H}$ has 
a $\Z$-grading induced by (\ref{E_Zgrading}) and by the grading of 
$\Omega^*(L)$ through form rank. This coincides with the 
grading (\ref{U1}) given by the worldsheet $U(1)$ charge, and whose 
mod $2$ reduction is the string field theoretic grading (\ref{eo}).

(2) One has boundary products, which in our case are given (up to signs) 
by the wedge product
of bundle valued forms (which are taken to involve composition 
of morphisms between the fibers):
\be
\label{products}
u\cdot v=(-1)^{(k-n)rank v} 
u\wedge v~~~~~~~{\rm for~}~u\in Hom(a_n,a_k)~{\rm~and~}~
v\in Hom(a_m,a_n)~~.
\ee
This gives compositions of the form 
$Hom(a_n,a_k)\times Hom(a_m,a_n)\rightarrow Hom(a_m,a_k)$. As in 
\cite{com1}, these induce a total boundary product on ${\cal H}$ through:
\be
u\cdot v=\oplus_{m,k}{\left[\sum_{n}{u_{nk}v_{mn}}\right]}
\ee
for elements $u=\oplus_{nk}{u_{nk}}$ and $v=\oplus_{mn}{v_{mn}}$ 
with $u_{nk}\in Hom(a_n, a_k)$ and $v_{mn}\in Hom(a_m, a_n)$.

The total boundary product also admits a standard supergeometric 
interpretation. Indeed, remember that both of the factors 
$\Omega^*(L)$ and $\Omega^0(End({\bf E}))$ admit natural structures 
of superalgebras, with multiplications given by wedge product of 
forms and composition of bundle morphisms, respectively. This 
allows us to consider the induced superalgebra structure on 
${\cal H}$, which, following \cite{Quillen}, corresponds to
$\Omega^*(L){\hat \otimes}_{\Omega^0(L)} \Omega^0(End({\bf E}))$.
According to standard supermathematics, the corresponding 
product on ${\cal H}$ acts on decomposable elements $u=\omega\otimes f$ 
and $v=\eta\otimes g$ as follows:
\be
(\omega\otimes f)(\eta\otimes g)=(-1)^{\pi(f) rank \eta}
(\omega\wedge \eta)\otimes (f\circ g)~~,
\ee
for $\omega, \eta$ some forms on $L$ and $f,g\in End({\bf E})$.
In this relation, $\pi(f)$ stands for the parity of $f$ 
with respect to the decomposition (\ref{superbundle}).

If we apply this relation for $u\in Hom(a_n, a_k)$ and $v\in 
Hom(a_m, a_n)$, then $\pi(f)$ is the mod $2$ reduction of $k-n$ and 
we recover relation (\ref{products}) upon viewing $u$ and $v$ 
as bundle-valued forms. In local coordinates, one can write write
$u=dx^{\alpha_1}\wedge ...\wedge dx^{\alpha_r}u_{\alpha_1..\alpha_r}$, 
$v=dx^{\alpha_1}\wedge ...\wedge dx^{\alpha_s}v_{\alpha_{1}...\alpha_s}$, 
and obtain:
\be
uv=(-1)^{s(k-n)}
dx^{\alpha_1}\wedge ...\wedge dx^{\alpha_{r+s}}u_{\alpha_1...\alpha_r}
v_{\alpha_{r+1}...\alpha_{r+s}}~~,
\ee
which recovers the boundary products introduced above. 
This means that one has a 
sign factor of $-1$ each time one commutes $dx^\alpha$ with an odd 
bundle morphism. 
In our conventions, one writes the bundle morphisms to the right.

(3) One has bilinear forms $_{mn}\langle . , . \rangle_{nm}$ 
on the products 
$Hom(a_m,a_n)\times Hom(a_n,a_m)$, which 
are induced by the two-point boundary correlator on the disk. 
These can be identified through a localization argument and are 
given by:
\be
\label{metrics}
_{mn}\langle u,v \rangle_{nm}=\int_{L_n}{tr_{E_n}(u\cdot v)}
=(-1)^n\int_{L}{tr_{E_n}(u\cdot v)}
\ee
where $tr_{E_n}$ denotes the fiberwise trace on $End(E_n)$. 

Since integration of forms requires the specification of an orientation, 
one must keep track of which of the two orientations of $L$ is used in the 
definition of each bilinear form. This gives the crucial sign factor in the 
right hand side.  As in \cite{com1}, one can combine these into a 
bilinear form $\langle . , .\rangle$ 
on the total boundary state space ${\cal H}$. Due to the 
sign factor in (\ref{metrics}), one obtains:
\be
\langle u , v \rangle=\int_L{str(uv)}~~{\rm for}~~u,v\in {\cal H}~~,
\ee
where $str$ is the supertrace with 
respect to the decomposition (\ref{superbundle}) (see \cite{Quillen} 
for details). 

(4) One has degree one BRST operators 
$Q_{mn}:Hom(a_m, a_n)\rightarrow Hom(a_m, a_n)$, which in our case are 
given by the covariant differentials $d_{mn}$ associated with the flat 
connections $A_m$ and $A_n$. More precisely, $d_{mn}$ is the 
differential of the `twisted' de Rham complex:
{\scriptsize\bea
0\stackrel{d_{mn}}{\longrightarrow}\Omega^{0}(L, Hom(E_m, E_n))
\stackrel{d_{mn}}{\longrightarrow}\Omega^1(L, Hom(E_m, E_n))
\stackrel{d_{mn}}{\longrightarrow}\Omega^2(L, Hom(E_m, E_n))
\stackrel{d_{mn}}{\longrightarrow}\Omega^3(L, Hom(E_m, E_n))
\stackrel{d_{mn}}{\longrightarrow} 0~~.\nn
\eea}
determined by the connection $\nabla_{mn}$ induced by 
$A_m$ and $A_n$ on $Hom(E_m, E_n)$.

These can be combined into the total BRST operator 
$Q=\oplus_{m,n}{Q_{mn}}$, which can be identified as the exterior 
differential on ${\bf E}$ induced by the total connection $A=\oplus_n{A_n}$. 
This is a differential operator on $\Omega^*(End({\bf E}))$ with the 
property:
\be
Q(\omega \otimes f)=d\omega \otimes f +(-1)^{rank \omega} \omega Qf~~,
\ee
so it gives a superconnection on the superbundle ${\bf E}$ in the 
sense of \cite{Quillen}. 

It is not hard to see that all of the axioms  discussed in \cite{com1} 
(and reviewed in Section 2) are satisfied. 
In the language of Section 2, we have 
a differential graded category ${\tilde {\cal A}}$ 
with objects $a_n$ and bilinear forms
between the morphism spaces which are invariant with respect to the 
BRST charges $Q_{mn}$ and with respect to morphism compositions.  
The abstract structure is depicted in figure 1.  

\hskip 1.0 in
\begin{center} 
\scalebox{0.6}{\begin{picture}(0,0)%
\includegraphics{cat.pstex}%
\end{picture}%
\setlength{\unitlength}{4144sp}%
\begingroup\makeatletter\ifx\SetFigFont\undefined%
\gdef\SetFigFont#1#2#3#4#5{%
  \reset@font\fontsize{#1}{#2pt}%
  \fontfamily{#3}\fontseries{#4}\fontshape{#5}%
  \selectfont}%
\fi\endgroup%
\begin{picture}(3410,2400)(387,-2446)
\put(406,-691){\makebox(0,0)[lb]{\smash{\SetFigFont{12}{14.4}{\familydefault}{\mddefault}{\updefault}
$Hom(a_m,a_m)$
}}}
\put(3331,-826){\makebox(0,0)[lb]{\smash{\SetFigFont{12}{14.4}{\familydefault}{\mddefault}{\updefault}
$Hom(a_n,a_n)$
}}}
\put(1846,-241){\makebox(0,0)[lb]{\smash{\SetFigFont{12}{14.4}{\familydefault}{\mddefault}{\updefault}
$Hom(a_m,a_n)$
}}}
\put(1756,-2446){\makebox(0,0)[lb]{\smash{\SetFigFont{12}{14.4}{\familydefault}{\mddefault}{\updefault}
$Hom(a_n,a_m)$
}}}
\put(946,-2041){\makebox(0,0)[lb]{\smash{\SetFigFont{12}{14.4}{\familydefault}{\mddefault}{\updefault}
$a_m$
}}}
\put(3196,-2131){\makebox(0,0)[lb]{\smash{\SetFigFont{12}{14.4}{\familydefault}{\mddefault}{\updefault}
$a_n$
}}}
\end{picture}
}
\end{center}
\begin{center} 
Figure  1. {\footnotesize 
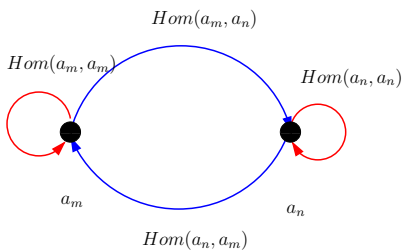
A full two-object subcategory of the category describing our D-brane system.
}
\end{center}

According to the axioms of Section 2, the open string field theory 
is described by the action:
\be
\label{sfa}
S(\phi)=\frac{1}{2}\langle \phi , Q\phi \rangle +\frac{1}{3}
\langle \phi, \phi\cdot \phi\rangle 
\ee
where the string field 
$\phi=\oplus_{m,n}{\phi_{mn}}$ (with $\phi_{mn}\in Hom(a_m, a_n)$) 
is a degree 
one
element of the total boundary state space 
${\cal H}$. 
Combining all of the data above, one can re-write this in 
the form:
\be
\label{action}
S(\phi)=\int_{L}{str
\left[\frac{1}{2}\phi Q\phi+\frac{1}{3}\phi\phi\phi\right]}~~.
\ee 
The product in the integrand is given by (\ref{products}).

\section{Complex conjugations}

Our string field theory can be endowed 
with conjugations, provided that the collection of D-branes 
$({\bf L}_n, E_n, A_n)$ is invariant with respect to the operation
which takes a brane into a conjugate brane\footnote{Our conjugate 
branes should {\em not} be identified with the `topological antibranes' of 
\cite{Douglas_Kontsevich, Aspinwall}. The `topological antibranes' of 
those papers result from our conjugate branes upon performing a further 
shift of the grading by $1$. The conjugation operators we construct below
are related with 
the `gauge invariance' of \cite{Douglas_Kontsevich, Aspinwall}, 
which consists in shifting the grading of all D-branes by $1$ combined 
with reversing the role of branes and antibranes.}. We propose that the  
conjugate brane of a D-brane $a=({\bf L}, E, A)$ is the D-brane  
${\overline a}=(\overline{\bf L}, {\overline E}, A^*)$ 
defined as follows: 

\

(a) The underlying cycle of $\overline{a}$ is $L$. 

\

(b) If the grading of $a$ is given by $n$ (i.e. ${\bf L}=(L, n)$), then 
the grading of ${\overline a}$ is ${\overline n}=-n$. 

\

(c) The underlying bundle ${\overline E}$ is the antidual of the bundle $E$, 
i.e. the bundle whose fiber $E_x$ at a point $x$ of $L$ is the space of 
{\em antilinear} functionals defined on the fiber $E_x$ of $E$. 

\

(d) The connection ${\overline \nabla}$ is given by:
\be
({\overline \nabla}_X\psi)(s)=-\psi(\nabla_{\overline X}(s))+X\left[\psi(s)\right]~~,
\ee
for any local sections $s$ of $E$ and $\psi$ of ${\overline E}$ and 
any vector fields $X$. In the right hand side, ${\overline X}$ is 
the complex conjugate of $X$. Namely, viewing $X$ as 
a complex-linear derivation of the algebra of (complex-valued) functions on $L$, one takes:
\be
{\overline X}(f):=\overline{X(\overline{f})}~~.
\ee
The result is another complex-linear derivation, 
i.e. a vector field. If $x^\alpha$
are (real !) local coordinates on $L$, then upon expanding $X=X^\alpha\partial_\alpha$, 
one has ${\overline X}=\overline{X^\alpha}\partial_\alpha$, 
where $\overline{X^\alpha}$ denotes 
the usual complex conjugate of the function $X^\alpha$. In particular, 
${\overline \partial_\alpha}=\partial_\alpha$.

If $s_\alpha$ is a local frame of sections for $E$, then 
we can consider the antidual frame ${\overline s}_\alpha$
of ${\overline E}$, defined by the conditions:
\be
{\overline s}_\alpha(s_\beta)=\delta_{\alpha\beta}~~.
\ee
Then we have (1-form valued) connection matrices $A$ and $A^*$ defined through:
\bea
\nabla(s_\alpha)&=&A_{\beta\alpha}s_\beta~~\\
{\overline \nabla}({\overline s}_\alpha)&=&
A^*_{\beta \alpha}{\overline s}_\beta~~.
\eea
In this case, we obtain the relation:
\be
A^*=-A^+~~, 
\ee
i.e.:
\be
A^*_i(x)=-A_i(x)^+~~,
\ee
where $A=A_i(x)dx^i$ and $A^*=A^*_i(x)dx^i$.

The abstract description given above has the advantage that it does 
not require the choice of supplementary data. It is possible to formulate 
a conjugation operator in these abstract terms, and check all relevant axioms 
given in Section 2. Below, I shall use a mathematically less elegant, 
but more concrete approach which requires the choice of a metric on 
the bundle $E$\footnote{This is 
not natural in a topological string theory, which should be formulated 
as much as possible without reference to a metric. I prefer to use 
the metric language in order to make the paper easier to understand 
for the casual reader.}. For this, let us assume that the bundle 
$E$ is endowed with a hermitian metric $h_E$; the 
precise choice of such a metric is irrelevant for what follows.
We use standard {\em physics} conventions by taking $h_E$ to be antilinear 
in its first variable and linear in the second. Such a scalar product 
defines a {\em linear} isomorphism between ${\overline E}$ and $E$, which 
identifies an antilinear functional $\psi$ on $E_x$ with 
the vector $u_\psi\in E_x$ satisfying the equation:
\be
h_E(v, u_\psi)=\psi(v)~~{\rm~for~all~}v\in E_x~~.
\ee
Using this isomorphism, one can translate the abstract definition 
of `conjugate branes' into the following more concrete description: 

\

(c') The underlying bundle ${\overline E}$ of ${\overline a}$ can be 
identified with the underlying bundle $E$ of $A$

\

(d') The background flat connection $A^*$ of $E$ can be identified 
with the opposite of the hermitian conjugate of $A$ with respect to the 
metric $h_E$:
\be
A^*=-A^+~~.
\ee
From now on, the symbol $A^+$ always denotes hermitian conjugation with 
respect to $h_E$, unless we state otherwise. We can now describe the 
antilinear conjugations of our string field theory. Since this is 
a bit subtle, we shall proceed in two steps. 

\paragraph{Step 1.} 
Given a (complex-valued) differential form $\omega=\omega_{\alpha_1..\alpha_k}
dx^{\alpha_1}\wedge ...\wedge dx^{\alpha_k}$ on $L$, we define
its conjugate by\footnote{$x^\alpha$ are {\em real} coordinates on the three-cycle 
$L$ (which is not a complex manifold !). I hope this avoids any confusion.}:
\be
\tilde{\omega}:=\overline{\omega_{\alpha_1...\alpha_k}}dx^{\alpha_k}\wedge 
...\wedge dx^{\alpha_1}=(-1)^{k(k-1)/2}{\overline \omega}~~,
\ee
where ${\overline \omega}=
\overline{\omega_{\alpha_1...\alpha_k}}dx^{\alpha_1}\wedge 
...\wedge dx^{\alpha_k}$ is the usual complex conjugate of $\omega$ and we 
used the relation:
\be
dx^{\alpha_k}\wedge ..\wedge dx^{\alpha_1}=(-1)^{k(k-1)/2}dx^{\alpha_1}
\wedge ...\wedge dx^{\alpha_k}~~.
\ee

It is not hard to check that this operation has the following properties:
\bea
\label{tilde}
(\alpha\wedge \beta)\tilde{~}&=&
\tilde{\beta}\wedge \tilde{\alpha}~~\nn\\
d(\tilde{\omega})&=&(-1)^{rank \omega}(d\omega)\tilde{~} ~~\\
\int_{L}{\tilde{\omega}}&=&-\overline{\int_{L}{\omega}}~~.
\eea

\paragraph{Step 2}

Let us now return to our our collection of D-branes 
$a_n=({\bf L}_n, E_n, A_n)$ 
and assume that we have picked hermitian metrics $h_n=h_{E_n}$ on 
each of the bundles $E_n$.
For $u=\omega\otimes f$ a decomposable element of 
$Hom(a_m, a_n)$ (with $\omega$ a complex-valued 
form on $L$ and $f\in Hom(E_m, E_n)$), we define:
\be
\label{conj}
*u:=(-1)^{(n-m+1)rk \omega}{\tilde \omega}\otimes f^+~~.
\ee 
We then extend this uniquely to an antilinear operation from 
$\Omega^*(L, Hom(E_m, E_n))$ to $\Omega^*(L, Hom(E_n, E_m))$.
In terms of the usual hermitian conjugation of bundle-valued forms
($(\omega\otimes f)^+:={\overline \omega}\otimes f^+$), this reads:
\be
*u=(-1)^{rk u (rk u+1)/2+(n-m)rk u }u^+~~.
\ee
Since the choice of hermitian metrics allows us to identify the 
`conjugate branes' 
${\overline a}_n=(\overline{L_n}, \overline{E_n}, A_n^*)$ 
with the triples $({\bf L}_{-n}, E_n, -A^+_n)$, we can also view 
(\ref{conj}) as antilinear maps $*_{mn}$ from $Hom(a_m, a_n)$ to 
$Hom(\overline{a_n}, \overline{a_m})$. We claim that these operators 
give conjugations of our string field theory. To see this, one must check 
that the axioms of Section 2 are satisfied. For this, notice first that 
the operations $*_{mn}$ are homogeneous of degree zero when viewed as 
applications between $Hom(a_m, a_n)$ and $Hom(\overline{a_n}, \overline
{a_m})$. This follows from the fact that they preserve form rank, and 
from form our definition ${\overline n}=-n$, ${\overline m}=-m$ 
of gradings for the `conjugate branes':
\be
|*_{mn}u_{mn}|=rank (*_{mn}u_{mn})+(-m)-(-n)=rank u_{mn}+n-m=|u_{mn}|~~.
\ee
The other properties listed in Subsection 2.2. then follow by straightforward 
computation, and I shall leave their verification as an exercise for the 
reader. 

It should be clear from our discussion that the theory 
considered in the previous section 
will be invariant under conjugation only 
if the set of D-branes $a_n$ is invariant with respect to the 
involution $a_n\rightarrow \overline{a_n}$. This can always be 
achieved by adding the `conjugate branes' $\overline{a_n}$ to the original 
set, but in general this will lead to more than one D-brane 
wrapped on $L$ for each grade $n$. Hence a general 
analysis requires that we allow various D-branes present in the system 
to have the same grade $n$. While it is possible to carry out 
our analysis for such general systems, this 
leads to rather complicated notation. 
When discussing reality issues in this paper, we shall 
assume for simplicity that the set of D-branes present in 
our background is invariant with respect to 
the transformation $a_n\rightarrow \overline{a_n}$. This requires that the set of 
grades under consideration is invariant with respect to the substitution 
$n\rightarrow -n$ and that $E_{-n}=E_n$, $h_{E_{-n}}=h_{E_n}$ and 
$A_{-n}=-A^+_n=A_n$. With these hypotheses, one 
has ${\overline a_n}=a_{-n}$, and we can impose the reality constraint 
$*\phi_{mn}=\phi_{-n, -m}$ on the string field. 
Note, however, that there is no reason why
our background should contain only brane-`conjugate brane' pairs. If this 
condition is not satisfied, then one can simply work with a 
{\em complex} action of the form (\ref{general_action}). 
In fact, complex string field actions are natural in topological
string field theories, as is well-known from the example of the B-model.
A real string field action is only required in a physically complete 
theory, such as the theory of `all' topological D-branes.

\section{An extended action}

Our  theory (\ref{action}) can be extended in the following manner. 
Consider the supermanifold
${\cal L}:=\Pi TL$ 
obtained by applying parity reversal on the fibers of the 
tangent bundle of $L$. ${\cal L}$ is equipped with a sheaf 
${\cal O}_{\cal L}$ of Grassmann algebras, whose sections are 
superfunctions defined on ${\cal L}$.

If $\theta^\alpha$ $(\alpha=1..3$) are odd coordinates
along the fibers $T_xL$, then sections of ${\cal O}_{\cal L}$ 
are superfields 
of the form:
\be
\label{taylor}
f(x, \theta)=f^{(0)}(x)+
\theta^ \alpha f^{(1)}_\alpha(x)+\theta^\alpha\theta^\beta
f^{(2)}_{\alpha\beta}(x)+\theta^\alpha\theta^\beta\theta^\gamma 
f^{(3)}_{\alpha\beta\gamma}(x)~~,
\ee
with Grassmann-valued coefficients $f^{(k)}_{\alpha_1..\alpha_k}$
(sections of ${\hat {\cal O}}_L={\cal O}_L\otimes G$, 
where $G$ is an underlying Grassmann 
algebra).
The space $\Gamma ({\cal O}_{\cal L})$ of such superfields forms a 
$\Z\times \Z_2$-graded 
algebra, with $\Z$-grading\footnote{The components 
of this grading are nonzero only in degrees $0,1,2$ and $3$.}
induced by the degree of 
the monomials in $\theta^\alpha$ and $\Z_2$-grading given by 
Grassmann parity.

We are interested in elements ${\bf \Phi}$ of the space
$\Gamma({\cal O}_{\cal L}{\hat \otimes}End({\bf E}))$, 
which carries the $\Z_2$ grading induced from the two components
of the tensor product. 
We shall denote this total $\Z_2$-grading by $deg$.
Such bundle-valued superfields have the expansion:
\be
\label{phi_expansion}
{\bf \Phi}(x,\theta)=\sum_{k=0}^3{\theta^{\alpha_1}..\theta^{\alpha_k}
{\bf \Phi}^{(k)}_{\alpha_1..\alpha_k}(x)}~~,
\ee
whose coefficients ${\bf \Phi}^{(k)}_{\alpha_1..\alpha_k}$ are sections of
the sheaf ${\bf End}({\bf E}):=
{\hat {\cal O}}_L\otimes End({\bf E})$, i.e. elements of the algebra
$\Gamma({\bf End}({\bf E})):=G{\hat \otimes}_\C \Gamma(L, End({\bf E}))$
of Grassmann-valued sections of $End({\bf E})$. This algebra has 
a $\Z$-grading induced by the $\Z$-grading on $End({\bf E})$ and a 
total $\Z_2$-grading induced by the sum of Grassmann parity with the 
$\Z_2$-degree on $End({\bf E})$. Note that the mod $2$ reduction of the 
$\Z$-grading does not give the $\Z_2$ grading.

The coefficients  ${\bf \Phi}^{(k)}_{\alpha_1..\alpha_k}$ 
define {\em Grassmann-valued} forms 
${\hat \phi}^{(k)}=
dx^{\alpha_1}\wedge ... 
\wedge dx^{\alpha_k}{\bf \Phi}^{(k)}_{\alpha_1..\alpha_k}(x)$, viewed as  
elements of the algebra 
${\cal H}_e=\Omega^*(L){\hat \otimes} 
\Gamma({\bf End}({\bf E}))$ of forms with Grassmann-valued coefficients in 
$End({\bf E})$. 
This algebra also has total 
$\Z$ and $\Z_2$-gradings (denoted again by $|.|$ and $deg$) 
which are induced by the $\Z$ and $\Z_2$-gradings on 
$\Omega^*(L)$ and $\Gamma({\bf End}({\bf E}))$. The $\Z_2$-grading 
$deg$ is the mod $2$ reduction of the 
sum between the $\Z$-grading $|.|$ and Grassmann parity.
If ${\hat a}$ and ${\hat b}$ are elements of ${\cal H}_e$, then 
their product is:
\be
\label{hat_product}
{\hat a}{\hat b}=
(-1)^{(deg {\hat a}-rank {\hat a})~rank {\hat b}}{\hat a}\wedge {\hat b}~~.
\ee

If $deg({\bf \Phi})=p\in \Z_2$ is the parity of the superfield ${\bf \Phi}$, 
then the components  ${\bf \Phi}^{(k)}_{\alpha_1..\alpha_k}$ 
have total parity 
$deg{\bf \Phi}^{(k)}_{\alpha_1..\alpha_k}=p-k (mod 2)$, 
while the Grassmann-valued form ${\hat \phi}^{(k)}$ has 
total parity $p$. 
We obtain a correspondence 
${\bf \Phi}\leftrightarrow {\hat \phi}$ which takes ${\bf \Phi}$ into
the sum of Grassmann-valued forms 
${\hat \phi}=\sum_{k=0}^3{{\bf \phi}^{(k)}}$. This map is homogeneous of 
degree zero, i.e. it intertwines the $\Z_2$ gradings $deg$ on 
$\Gamma({\cal O}_{\cal L}\otimes End({\bf E}))$ and ${\cal H}_e$.

Under this correspondence, the BRST differential 
$Q=d$ on ${\cal H}_e$ 
(extended to Grassmann-valued forms in the obvious manner) 
maps to the operator:
\be
D=\theta^\alpha\frac{\partial}{\partial x^\alpha}~~
\ee
on superfields.
By analogy with the usual Chern-Simons case, this 
allows us to write an extended string field action:
\be
\label{extended_action}
S_e({\bf \Phi})=
\int_L d^3x\int{d^3\theta 
str\left[\frac{1}{2}{\bf \Phi} {\bf D}{\bf \Phi}+
\frac{1}{3}{\bf \Phi}{\bf \Phi}{\bf \Phi}
\right]}~~.
\ee
This can be viewed as a $\Z$-graded version of {\em extended} 
super-Chern-Simons field theory. It should be compared with the proposal of 
\cite{Vafa_cs}.

\paragraph{Observation}

It is a standard subtlety (see, for example, \cite{AS1}) that the product 
of the Grassmann-valued forms 
${\hat \phi^{(k)}}$ induced by superfield multiplication 
differs from the standard wedge product. 
This is due to the fact that the coordinates 
$\theta^\alpha$ are Grassmann odd. To be precise, let us consider 
two superfields ${\bf A}$, ${\bf B}$ 
of parities $p_{\bf A}$ and $p_{\bf B}$, and let 
${\bf A}^{(k)}_{\alpha_1..\alpha_k}$ and 
${\bf B}^{(l)}_{\alpha_1..\alpha_l}$ 
be their components under the expansion (\ref{taylor}), 
which have  total 
parities $p_{\bf A}-k (mod 2)$ and $p_{\bf B}-l (mod 2)$.      
We also consider the associated Grassmann-valued forms ${\hat a}^{(k)}=
dx^{\alpha_1}\wedge ...\wedge dx^{\alpha_k}{\bf A}^{(k)}_{\alpha_1..\alpha_k}$ 
and ${\hat b}^{(l)}=
dx^{\alpha_1}\wedge ...\wedge dx^{\alpha_l}{\bf B}^{(l)}_{\alpha_1..\alpha_l}$.
Then one can write:
\be
({\bf A}{\bf B})(x,\theta)=
\sum_{n=0}^3{\theta^{\alpha_1}...\theta^{\alpha_n}
({\bf A}{\bf B})^{(n)}_{\alpha_1...\alpha_n}}(x)~~.
\ee
If $({\hat a}*{\hat b})^{(n)}=
dx^{\alpha_1}\wedge ...\wedge dx^{\alpha_n}
({\bf A}{\bf B})^{(n)}_{\alpha_1..\alpha_n}(x)$ 
is the associated Grassmann-valued form, then one has
$({\hat a}*{\hat b})^{(n)}=\sum_{k+l=n}{{\hat a}^{(k)}*{\hat b}^{(l)}}$, where:
\be
\label{sf_star}
{\hat a}^{(k)}*{\hat b}^{(l)}=(-1)^{(p_{\bf A}-k)l}{\hat a}^{(k)}
\wedge {\hat b}^{(l)}~~.
\ee
The sign prefactor arises 
when commuting the odd variables $\theta^{\beta_j}$ with the coefficient
${\bf A}^{(k)}_{\alpha_1..\alpha_k}$ in the following relation:
\be
\theta^{\alpha_1}..\theta^{\alpha_k} {\bf A}^{(k)}_{\alpha_1..\alpha_k}
\theta^{\beta_1}...\theta^{\beta_l}{\bf B}^{(l)}_{\beta_1..\beta_l} =
(-1)^{(p_{\bf A}-k)l}\theta^{\alpha_1}..\theta^{\alpha_k}
\theta^{\beta_1}...\theta^{\beta_l}{\bf A}^{(k)}_{\alpha_1..\alpha_k}
{\bf B}^{(l)}_{\beta_1..\beta_l} 
\ee
Equation (\ref{sf_star}) reads:
\be
\label{star}
{\hat a}*{\hat b}=(-1)^{(p_{\bf A}-rank{\hat a})rank {\hat b}
}{\hat a}\wedge {\hat b}=(-1)^{(p_{\bf A}-|{\bf a}|)rank {\hat b}
}{\hat a}\cdot {\hat b}~~,
\ee 
where the multiplication $\cdot$ is that of (\ref{products}). 
Since $p_{\bf A}=deg {\hat a}$, 
this is exactly the product (\ref{hat_product}) on the algebra ${\cal H}_e$. 
Hence one can write the extended action as a functional on ${\cal H}_e$:
\be
S_e({\hat \phi})=
\int_L{d^3x 
str\left[\frac{1}{2}{\hat \phi} *d{\hat \phi}+
\frac{1}{3}{\hat \phi}*{\hat \phi}*{\hat \phi}
\right]}~~.
\ee
In this form, the action $S_e$ can be related to the general extension 
procedure discussed in \cite{Gaberdiel} (when the latter is 
translated in our conventions).
One imposes the condition that ${\bf \Phi}$ is an odd superfield, i.e. 
$deg {\hat \phi}={\hat 1}\in \Z_2$. 

If $p_{\bf A}={\hat 1}$, 
then the product (\ref{star}) agrees with the multiplication 
(\ref{products}) when $|{\bf a}|=1$. This implies that the extended 
action (\ref{extended_action}) 
reduces to the unextended one 
(\ref{sfa}) if all non-vanishing Grassmann-valued form 
components 
of the extended string field ${\bf \Phi}$ satisfy 
$|{\hat \phi}_{mn}|=1$.

\section{The moduli space of vacua}

Recall that the moduli space of vacua is built by solving the Maurer-Cartan 
equation:
\be
\label{mc}
Q\phi+\frac{1}{2}\left[\phi, \phi \right]=0~~\Leftrightarrow 
Q\phi+\phi\phi=0
\ee
for string fields $\phi\in {\cal H}$ of degree $|\phi|=1$. 
Two solutions of (\ref{mc}) are identified if they differ by 
a gauge transformation. The gauge group is generated 
by infinitesimal transformations of the form 
$\phi\rightarrow \phi+Qu+\left[\phi,u\right]$. 
If our collection of D-branes is invariant 
under complex conjugation, one can also impose 
the reality constraint $*\phi_{m,n}=\phi_{-n, -m}$, in which case 
one also requires reality ($*u=u$) of the gauge generator $u$. 
The commutator appearing in (\ref{mc}) is 
the graded commutator on the algebra ${\cal H}$, which reduces to the 
square of $\phi$ due to the degreee one condition $|\phi|=1$. 
The resulting moduli space is a (formal) supermanifold ${\cal M}$, which 
can be built 
locally by solving (\ref{mc}) upon expressing $\phi$ as a formal power 
series in appropriate variables \cite{Barannikov_formality}.

The degree one condition on the string field reads:
\be
\label{deg}
rank\phi_{mn}+n-m=1\Leftrightarrow rank\phi_{mn}=1+m-n~~. 
\ee
As in \cite{com1}, this implies that one can deform the vacuum by 
condensing fields of rank $1+m-n$  in the boundary sector $Hom(a_m,a_n)$. 
The fact that one can condense higher rank forms in this manner is due 
to the presence of $\Z$-graded branes in the background. In fact, it 
can be argued that the standard deformations (\ref{mc}) represent 
{\em extended} deformations of the usual vacuum based on the 
collection of flat connections $A_n$.

\subsection{The shift-invariant case}

To understand this, let us consider the case $E_m=E$ and $A_m=A$  
for all $m$, i.e. we take the underlying bundles and flat connections 
to be identical.
We also take an infinity of $D$-branes, i.e. we let $n$
run over all integer values from $-\infty$ to $+\infty$.
In this case, the moduli space built by solving (\ref{mc}) will in 
fact be ill-defined, due to the existence of a countable 
number of shift symmetries 
$S_k:\phi_{m,n}\rightarrow \phi_{m+k, n+k}$ of the string field action.

This problem is easily solved by restricting to
`shift-invariant' configurations, i.e. configurations of the string field 
for which $\phi_{mn}=\phi_{m-n}$ depends only on the difference $m-n$
\footnote{ A more careful treatment requires an analysis of 
shift-equivariant string field configurations, i.e. configurations which 
are shift invariant only up to gauge transformations. Here we give a 
simplified discussion in terms of shift-invariant solutions.}. 
We thus replace ${\cal M}$ by the moduli space ${\cal M}_d$, which is 
obtained by solving the Maurer-Cartan equations (\ref{mc}) for shift
invariant configurations. For such configurations, the equations reduce to:
\be
\label{mc_reduced}
Q_{n}\phi_n+\sum_{k+l=n}{\phi_k\phi_l}=0~~,
\ee
where $Q_n:=d_A$ acts on 
$Hom(a_n,a_0)=\Omega^*(L, End(E))[-n]$ and 
$\phi_n$ is a degree one element of  
$Hom(a_n, a_0)$, i.e. a form of rank $1+n$ valued 
in $End(E)$.  

We next consider the linearization $Q_n\phi_n=0$ of
(\ref{mc_reduced}), which describes infinitesimal 
vacuum  deformations. Upon dividing through linearized gauge 
transformations $(\phi_n\rightarrow \phi_n+Q_nu_n)$, 
we obtain that the tangent 
space to ${\cal M}_d$ (at the point $O=\{A_n=A~{\rm~for~all~}n\}$) 
is given by the degree one BRST cohomology of 
${\cal H}_d:=\oplus_{n}{\Omega^*(L, End(E))[-n]}$
\be
T_O{\cal M}_d=H_Q^1({\cal H}_d)=
ker \left[Q:{\cal H}^1_d\rightarrow {\cal H}^2_d\right]/
im \left[Q:{\cal H}^0_d\rightarrow {\cal H}^1_d\right]~~.
\ee
It is easy to see that this coincides with the {\em total} BRST cohomology 
of $\Omega^*(L, End(E))$:
\be
T_O({\cal M}_d)=H^*_{d_A}(L, End(E))=\oplus_{k=0}^3{H^k_{d_A}(L, End(E))}~~.
\ee

That is, (shift-invariant) {\em degree one deformations of our 
theory correspond to extended deformations of the moduli space of flat 
connections on $E$}. 
It is well-known (see, for example, \cite{Witten_CS}) that 
the moduli space of flat connections on $E$ is the same as the moduli space 
of (classical) vacua of the Chern-Simons theory based on $E$, which in 
turn gives the (large radius) description of the boundary sector of 
a {\em single} D-brane $({\bf L}_0, E, A)$ wrapped on $L$.
It follows that inclusion of graded D-branes leads automatically to 
an {\em extended} moduli space. This is a particular realization of 
the general principle (already mentioned in \cite{com3}) that 
usual 
(degree one) deformations of a shift-completed string field theory describe
{\em extended} deformations of the uncompleted system. 

Let us compare this with a theory of non-graded branes.
If one were to neglect 
the fact that type A branes are $\Z$-graded, and consider the naive 
identification of D-branes with the data $(L, E, A)$, then, as in 
\cite{Witten_CS}, one would arrive at the conclusion that the 
string field theory of the D-brane collection $(a_n)_n$ is the standard 
Chern-Simons field theory based on the bundle $E_{tot}=E_n^{\oplus \Z}$
(viewed as an {\em even} bundle rather than a superbundle) and 
considered around the vacuum described by the  
configuration $\{A_n=A~{\rm~for~all~}n\}$. 
Vacuum deformations of this theory would 
correspond to independent deformations $\phi_{n,n}$
of the flat connections $A_n$, 
and would locally give an infinite power ${\cal M}={\cal M}_0^{\Z}$ 
of the moduli space ${\cal M}_0$ of a single flat connection 
$A$. Restricting to shift-invariant deformations would then give the 
`regularized' moduli space ${\cal M}_d={\cal M}_0$. 
Tangent direction to the latter are described by bundle-valued 
{\em one}-forms $\omega\in \Omega^1(L, End(E))$.

\subsection{Degree constraints}

Let us now return to the general case of distinct bundles $E_n$. 
We would like to use constraint (\ref{deg}) in order to 
understand the types of 
D-brane configurations which result from physical vacuum deformations. 

For this, note that presence of the D-branes 
$a_m$ and $a_n$ in the string background requires consideration of 
both $\phi_{mn}$ and $\phi_{nm}$ as components of the string field. 
Since one requires that the total string field have degree one, then 
one obtains constraints in the case 
$\phi_{nm}\neq 0$ or $\phi_{mn}\neq 0$, 
due to the fact that the ranks of these components must belong to the 
set $\{0,1,2,3\}$. In fact, {\em both} of these components will vanish 
unless $|m-n|\leq 2$, i.e. $m-n\in \{-2,-1,0,1,2\}$. Since (\ref{deg})
simply requires $rank \phi_{mm}=1$ for $m=n$, we have {\em six} 
possibilities for which at least one of $\phi_{mn}$ and $\phi_{nm}$ can 
be nonzero:

\

(1)$|m-n|\leq 2$~~, which gives 5 possibilities, further 
subdivided as follows:

\

(a)$|m-n|=2$, i.e. $n\in \{m-2, m+2\}$, in which case only one of 
$\phi_{mn}$ and $\phi_{nm}$ can be nonzero

(b)$|m-n|=1$, i.e. $n\in \{m-1,m+1\}$, in which case both $\phi_{mn}$ 
and $\phi_{nm}$ can be nonzero.

(c)$m=n$, in which case $\phi_{mn}=\phi_{nm}=\phi_{mm}$ can be nonzero.

\

(2)$|m-n|>2$, which requires $\phi_{mn}=\phi_{nm}=0$. 

\

This can be visualized in the following manner. 
Let us identify the D-branes $a_n$ as 
abstract points of a regular one-dimensional lattice, and view the components 
$\phi_{mn}$ as link variables connecting the various nodes, with $\phi_{mm}$ 
viewed as self-linking of a node with itself. The links $\phi_{mn}$ for 
$m\neq n$ are oriented (since $\phi_{mn}$ and $\phi_{nm}$ should be 
viewed as independent data), while the self-links $\phi_{mm}$ carry no 
orientation. 

\hskip 1.0 in
\begin{center} 
\scalebox{0.8}{\begin{picture}(0,0)%
\includegraphics{links.pstex}%
\end{picture}%
\setlength{\unitlength}{4144sp}%
\begingroup\makeatletter\ifx\SetFigFont\undefined%
\gdef\SetFigFont#1#2#3#4#5{%
  \reset@font\fontsize{#1}{#2pt}%
  \fontfamily{#3}\fontseries{#4}\fontshape{#5}%
  \selectfont}%
\fi\endgroup%
\begin{picture}(6473,2280)(927,-2311)
\put(4591,-556){\makebox(0,0)[lb]{\smash{\SetFigFont{17}{20.4}{\familydefault}{\mddefault}{\updefault}
(0)
}}}
\put(3961,-2311){\makebox(0,0)[lb]{\smash{\SetFigFont{17}{20.4}{\familydefault}{\mddefault}{\updefault}
$m$
}}}
\put(5131,-2311){\makebox(0,0)[lb]{\smash{\SetFigFont{17}{20.4}{\familydefault}{\mddefault}{\updefault}
$m+1$
}}}
\put(2791,-2311){\makebox(0,0)[lb]{\smash{\SetFigFont{17}{20.4}{\familydefault}{\mddefault}{\updefault}
$m-1$
}}}
\put(1891,-2311){\makebox(0,0)[lb]{\smash{\SetFigFont{17}{20.4}{\familydefault}{\mddefault}{\updefault}
$m-2$
}}}
\put(3916,-241){\makebox(0,0)[lb]{\smash{\SetFigFont{17}{20.4}{\familydefault}{\mddefault}{\updefault}
(1)
}}}
\put(3466,-601){\makebox(0,0)[lb]{\smash{\SetFigFont{17}{20.4}{\familydefault}{\mddefault}{\updefault}
(2)
}}}
\put(2701,-421){\makebox(0,0)[lb]{\smash{\SetFigFont{17}{20.4}{\familydefault}{\mddefault}{\updefault}
(3)
}}}
\put(6166,-2266){\makebox(0,0)[lb]{\smash{\SetFigFont{17}{20.4}{\familydefault}{\mddefault}{\updefault}
$m+2$
}}}
\end{picture}
}
\end{center}
\begin{center} 
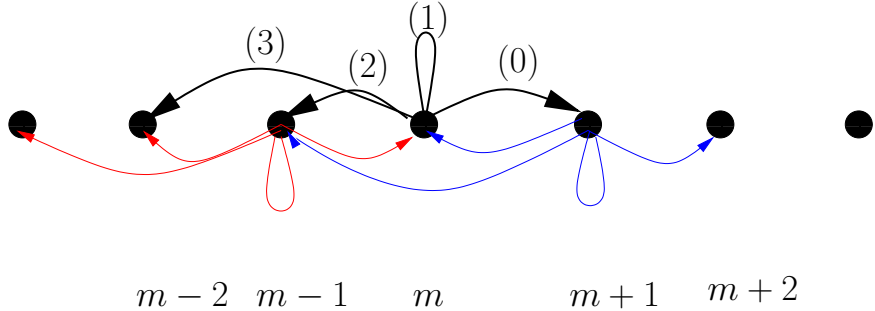
Figure  2. {\footnotesize The formal lattice describing the allowed 
string field configurations. The numbers in round brackets indicate 
form degree.}
\end{center}

Then $(1)$ and $(2)$ tell us that two nodes $m$ and $n$ can be 
connected by a 
link if and only if $|m-n|\leq 2$. Physically, this means that 
condensation 
of string field components is {\em local} with respect to the grade $n$, 
i.e. our system behaves in certain ways 
like a lattice with finite length interactions.  
This observation gives a 
string field theoretic interpretation of the point made in 
\cite{Douglas_Kontsevich} that the grade should in a certain sense 
be `$\Z_6$-valued'. In the conjugation-invariant case, 
the reality condition 
$*\phi_{mn}=\phi_{-n,-m}$ puts further constraints on the allowed 
configurations, without modifying the qualitative picture discussed 
above.

\section{Analysis of deformations}

Consider the general expansion of a degree one string field:
\be
\phi=\oplus_{mn}{\phi_{mn}}~~,
\ee
with $\phi_{mn}\in Hom^1(a_m, a_n)=\Omega^{1+m-n}(L, Hom(E_m, E_n))$. 

Only terms with $n-m=-2,-1,0$ or $1$ survive, so we obtain:
\be
\phi=\oplus_{m}{
\left[\phi^{(3)}_{m,m-2}\oplus\phi^{(2)}_{m,m-1}\oplus \phi^{(1)}_{m,m}\oplus 
\phi^{(0)}_{m,m+1}\right]}~~,
\ee
where the superscripts indicate the rank of forms.
The Maurer-Cartan equations read:

\be
d\phi^{(1+m-k)}_{mk}+\sum_{n}{\phi^{(1+n-k)}_{nk}\phi^{(1+m-n)}_{mn}}=0~~.
\ee
Considering the nontrivial cases $k=m-1, m, m+1, m+2$, this gives:
{\footnotesize \bea
\label{eq}
d\phi^{(2)}_{m,m-1}+\phi^{(0)}_{m-2,m-1}\phi^{(3)}_{m,m-2}
+\phi^{(1)}_{m-1,m-1}\phi^{(2)}_{m,m-1}+\phi^{(2)}_{m,m-1}\phi_{m,m}^{(1)}+
\phi^{(3)}_{m+1,m-1}\phi^{(0)}_{m,m+1}&=&0~~\nn~~~~~~~~~~~\\
d\phi^{(1)}_{m,m}+\phi^{(0)}_{m-1,m}\phi^{(2)}_{m,m-1}+
\phi_{m,m}^{(1)}\phi^{(1)}_{m,m}+\phi^{(2)}_{m+1,m}\phi^{(0)}_{m,m+1}&=&0~~,\nn\\
d\phi^{(0)}_{m,m+1}+\phi^{(0)}_{m,m+1}\phi_{m,m}^{(1)}+
\phi_{m+1,m+1}^{(1)}\phi_{m,m+1}^{(0)}&=&0~~\\
\phi^{(0)}_{m+1,m+2}\phi^{(0)}_{m,m+1}&=&0~~,\nn    
\eea}\noindent We obtain 4 systems of equations $\Sigma_k$ $(k=1..4$), where 
$\Sigma_k$ correspond to the $k$-th row in (\ref{eq}). 
Each system contains a number of equations parameterized by $m$.
In a theory with conjugations, 
each of the systems $\Sigma_k$ is
independently conjugation invariant, as a consequence of 
the reality condition $*\phi_{mn}=\phi_{-n,-m}$ on the string field.
These equations are a particular realization of similar 
constraints holding in an arbitrary cubic string field theory with 
D-branes, a more systematic study of which will be given in \cite{us}. 
Here I shall only make a few basic observations.

\subsection{Diagonal deformations} 

Let us consider the particular case $\phi_{mn}=\delta_{mn}
\phi^{(1)}_{mm}$, which corresponds to condensing boundary operators 
in each diagonal sector, but no boundary condition changing operators. 
In this situation, equations (\ref{eq}) reduce to:
\be
d\phi^{(1)}_{mm}+\phi^{(1)}_{mm}\phi^{(1)}_{mm}=0~~,
\ee
with 
$d=d_{mm}$, the differential on $\Omega^*(L, End(E_m))$ 
induced by the flat connection $A_m$.
Since $\phi^{(1)}_{mm}$ are one-forms valued in $End(E_m)$, and 
the product (\ref{products}) reduces in this case to the 
usual wedge product of forms, this can also be written as:
\be
d\phi^{(1)}_{mm}+\phi^{(1)}_{mm}\wedge \phi^{(1)}_{mm}=0~~.
\ee
These are the standard equations describing {\em independent} 
deformations 
$A_m\rightarrow A'_m:=A_m+{\phi^{(1)}}_{mm}$ of the flat connections $A_m$. 
This is a realization of the general principle \cite{com1, com3} 
that condensation of 
diagonal components of the string field can be interpreted as 
performing independent deformations of the D-branes $a_n$.
Upon integrating such deformations, we obtain a D-brane system of 
the type ${(a'_n)}_n$, where $a'_n$ is a deformation of the brane $a_n$, 
obtained by modifying its flat background connection. 
More precisely, one obtains $a'_n=({\bf L}_n, E_n, A'_n)$, 
with the new background connection $A'_n$. In the language 
of \cite{com1, com3}, such deformations preserve the category structure 
and thus they do not lead to D-brane composite formation.

\subsection{Unidirectional off-diagonal deformations and exotic type A 
branes}

Let us now consider the case where all $\phi_{mm-2}$, $\phi_{mm-1}$
and $\phi_{mm}$ vanish  
vanish but $\phi_{mm+1}$ may be non-zero. This corresponds to 
condensing degree zero forms $\phi^{(0)}_{mm+1}$ in each 
boundary condition changing sector $Hom(a_m, a_{m+1})$. 
Note that $\phi^{(0)}_{mm+1}$ is simply a bundle 
morphism from $E_m$ to $E_{m+1}$.

In this case, the 
constraints (\ref{eq}) reduce to:
\bea
\phi^{(0)}_{m+1,m+2}\phi^{(0)}_{m,m+1}&=&0~~\\    
d\phi^{(0)}_{m,m+1}&=&0~~.
\eea
The second condition tells us that the morphism $\phi_{mm+1}$ is 
flat (covariantly constant) 
with respect to the connection induced by $A_m$ and $A_{m+1}$ 
on the bundle $Hom(E_m, E_{m+1})$, while the first equation shows 
that these morphisms form a {\em complex}:
\be
\label{complex}
...~E_{m-1}\stackrel{\phi_{m-1,m}}{\longrightarrow}
E_m\stackrel{\phi_{m, m+1}}{\longrightarrow}E_{m+1}~...~~.
\ee

As explained on general grounds in \cite{com1, com3}, condensation 
of such operators destroys the original category structure, leading to 
a so-called {\em collapsed category}. If we restrict to the simplest 
case when the collection of nonzero condensates $\phi_{mm+1}$ connects 
together all of the branes present in our system\footnote{That is, 
if none of the morphisms in the complex (\ref{complex}) vanishes.}, 
then the end 
result of such a condensation process is a single D-brane composite. 
It follows 
that one can produce an {\em   entirely new object} associated with the 
cycle $L$ through such a process. Such an object corresponds to a 
complex of the type (\ref{complex}), whose morphisms are covariantly 
constant with respect to the original connections $A_n$.  
Complexes of this type were studied from a mathematical perspective 
in \cite{Bismut_Lott}. We wish to stress that the entire complex 
(\ref{complex}) must be viewed as a {\em new} topological D-brane 
(a D-brane composite) in this situation. In the case when some of the 
morphisms $\phi_{m,m+1}$ vanish, the complex (\ref{complex}) splits 
into connected subcomplexes, and each subcomplex should be viewed as 
a novel D-brane composite.

We conclude that, 
{\em given a special Lagrangian cycle $L$, one has many more 
topological D-branes wrapping $L$ 
than the graded branes of the type $({\bf L}_n, E_n,A_n)$}. 
This already gives a vast enlargement of the category of topological 
D-branes, as was already mentioned in \cite{com3}. Of course, an even 
bigger enlargement can be obtained by considering more than one 
Lagrangian cycle, for example by taking a collection of Lagrangians
with transverse intersections. This will be discussed in detail 
somewhere else. 

\subsection{General deformations}

It is clear that a generic deformation satisfying (\ref{eq}) does 
not correspond to a flat connection background on the cycle $L$; 
indeed, such deformations involve condensation of forms of rank 
{\em zero}, {\em two} and {\em three}, beyond the standard condensation 
of rank one forms.  
The resulting string field backgrounds therefore 
lead to quite exotic classes of topological D-brane composites. 
The formalism appropriate for 
studying such systems is the theory of flat superconnections 
on $\Z$-graded superbundles, the basics of which were developed by Bismutt 
and Lott in \cite{Bismut_Lott}. This must be combined with the foundational 
work of \cite{com1, com3} and with the mathematical discussion of 
Bondal and Kapranov \cite{BK}. 

\subsubsection{General string field backgrounds and pseudocomplexes}

While I will not give a complete analysis 
along these lines, I wish to explain the relation between 
general solutions to (\ref{eq}) and the framework of 
\cite{com1, com3}. For this, let us once again restrict to the 
shift-invariant case $E_n=E$ and $A_n=A$ for all $n$, and let 
us assume that $n$ runs over all signed integers.

In the language of \cite{com3},
our category ${\tilde {\cal A}}$ built on the objects $a_n$ is then the 
{\em shift completion} of the one-object category ${\cal A}$ formed 
by the D-brane $a_0=({\bf L}_0, E, A)$ together with the morphism 
space $Hom(a_0, a_0)$ and the induced morphism composition (figure 2).
Indeed, the objects $a_n$ can be identified with the {\em formal translates} 
$a_n=a[n]$ of the object $a_0:=a$, and the morphism spaces 
$Hom(a_m, a_n)=Hom(a[m],a[n])$ are given by:
\be
Hom(a[m], a[n])=Hom(a,a)[n-m]~~.
\ee
 
The one-object category ${\cal A}$ 
describes the boundary sector of a 
single D-brane $a_0$, and underlies 
the open string field theory of \cite{Witten_CS}, which 
is equivalent with the standard Chern-Simons field theory on the 
bundle $E_0=E$. 
Indeed, since $m=n=0$, the extra signs in the boundary product and 
topological metric of Section 3 dissapear in this case. 
Hence {\em inclusion of graded D-branes $a_n$ amounts to taking 
the shift completion ${\tilde {\cal A}}$ 
of the naive one-objects category ${\cal A}$}. That is, 
{\em working with graded D-branes amounts to taking the shift-completion}.

This is in fact a 
general principle, already pointed out in \cite{com3}, 
which gives a more conceptual explanation for 
the observations of \cite{Douglas_Kontsevich} and whose origin 
can be traced back to the theory of BV quantization. It is also 
intimately related with extended deformation theory
\cite{Kontsevich_Felder, Manin}, as will be discussed in detail in 
\cite{us}.

\hskip 1.0 in
\begin{center} 
\scalebox{0.8}{\begin{picture}(0,0)%
\includegraphics{one_object.pstex}%
\end{picture}%
\setlength{\unitlength}{4144sp}%
\begingroup\makeatletter\ifx\SetFigFont\undefined%
\gdef\SetFigFont#1#2#3#4#5{%
  \reset@font\fontsize{#1}{#2pt}%
  \fontfamily{#3}\fontseries{#4}\fontshape{#5}%
  \selectfont}%
\fi\endgroup%
\begin{picture}(772,1506)(1621,-1501)
\put(1936,-1501){\makebox(0,0)[lb]{\smash{\SetFigFont{12}{14.4}{\familydefault}{\mddefault}{\updefault}
$a_0$
}}}
\put(1621,-151){\makebox(0,0)[lb]{\smash{\SetFigFont{12}{14.4}{\familydefault}{\mddefault}{\updefault}
$Hom(a_0,a_0)=\Omega^*(L, End(E)))$
}}}
\end{picture}
}
\end{center}
\begin{center} 
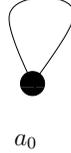
Figure  3. {The open string field theory 
of \cite{Witten_CS} 
(identical with standard Chern-Simons field theory on $L$) 
corresponds to a one-object category ${\cal A}$, whose shift completion 
${\tilde {\cal A}}$ 
gives the string field theory of the present paper, if one restricts to the 
shift-invariant case $E_n=E_0$ and $A_n=A_0$.}
\end{center}

In this categorical language, 
general solutions of (\ref{eq}) correspond to so-called 
{\em pseudocomplexes} \cite{com1} built out of objects and morphisms 
of the shift-completed category ${\tilde {\cal A}}$. For example, 
a solution of (\ref{eq}) containing only four nonvanishing 
components $\phi_{m,m-1}$, $\phi_{m,m}$, 
$\phi_{mm+1}$ and $\phi_{m+1,m-1}$ 
(for some fixed $m$) corresponds to the pseudocomplex
depicted in figure 4.

\hskip 1.0 in
\begin{center} 
\scalebox{0.8}{\begin{picture}(0,0)%
\includegraphics{pseudocomplex.pstex}%
\end{picture}%
\setlength{\unitlength}{4144sp}%
\begingroup\makeatletter\ifx\SetFigFont\undefined%
\gdef\SetFigFont#1#2#3#4#5{%
  \reset@font\fontsize{#1}{#2pt}%
  \fontfamily{#3}\fontseries{#4}\fontshape{#5}%
  \selectfont}%
\fi\endgroup%
\begin{picture}(3158,1902)(136,-1501)
\put(136,-331){\makebox(0,0)[lb]{\smash{\SetFigFont{17}{20.4}{\familydefault}{\mddefault}{\updefault}
$\phi^{(2)}_{m,m-1}$
}}}
\put(2566,-331){\makebox(0,0)[lb]{\smash{\SetFigFont{17}{20.4}{\familydefault}{\mddefault}{\updefault}
$\phi^{(0)}_{m,m+1}$
}}}
\put(1666,209){\makebox(0,0)[lb]{\smash{\SetFigFont{17}{20.4}{\familydefault}{\mddefault}{\updefault}
$\phi^{(1)}_{mm}$
}}}
\put(1621,-1141){\makebox(0,0)[lb]{\smash{\SetFigFont{17}{20.4}{\familydefault}{\mddefault}{\updefault}
$\phi^{(3)}_{m+1,m-1}$
}}}
\put(3016,-1501){\makebox(0,0)[lb]{\smash{\SetFigFont{17}{20.4}{\familydefault}{\mddefault}{\updefault}
$a_{m+1}$
}}}
\put(1846,-1501){\makebox(0,0)[lb]{\smash{\SetFigFont{17}{20.4}{\familydefault}{\mddefault}{\updefault}
$a_m$
}}}
\put(541,-1501){\makebox(0,0)[lb]{\smash{\SetFigFont{17}{20.4}{\familydefault}{\mddefault}{\updefault}
$a_{m-1}$
}}}
\end{picture}
}
\end{center}
\begin{center} 
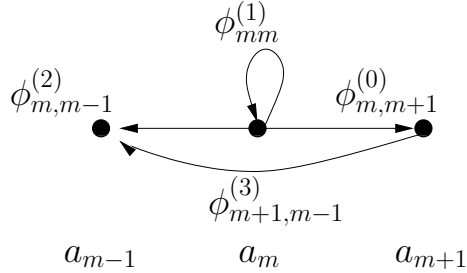
Figure  4. {\footnotesize A pseudocomplex formed by three objects $a_{m-1}$, 
$a_m$ and $a_{m+1}$ and four morphisms $\phi_{m,m-1}$, 
$\phi_{m,m}$, $\phi_{m,m+1}$ and $\phi_{m+1,m-1}$ of ${\tilde {\cal A}}$.} 
\end{center}

It was shown on general grounds in 
\cite{com1, com3} that pseudocomplexes form a dG category 
$p({\tilde {\cal A}})$ of their own 
and, in fact, give an admissible class of D-branes which extends the class 
of objects of ${\tilde {\cal A}}$. 
This allows for a string field theoretic description of any D-brane 
configuration resulting from condensation of string field components 
$\phi_{mn}$ satisfying equations (\ref{eq}). The resulting string field 
theory gives a description of open string dynamics in 
the presence of such general condensates, which include the original D-branes
$a_n$, covariantly constant complexes of the type 
(\ref{complex}) as well as much 
more general objects, which do not admit a classical geometric description. 
{\em It is this category of `exotic A-type branes', rather than the naive 
one-object category 
${\cal A}$ which must be studied in order 
to gain a better understanding of open string mirror symmetry `with one cycle'}.
This gives a very nontrivial extension of the theory originally considered 
in \cite{Witten_CS}, and provides a first step toward an open string 
realization of the program outlined 
in \cite{Witten_mirror} of gaining a better understanding of mirror symmetry 
by considering the {\em extended} moduli space of topological strings. 

\subsubsection{Relation with work of Bondal and Kapranov}
As pointed out in \cite{com1, com3}, pseudocomplexes over ${\tilde {\cal A}}$ 
allow one to make 
contact with the so-called {\em enhanced triangulated categories} 
discussed by Bondal and Kapranov \cite{BK}. In fact, it is 
easy to see that pseudocomplexes over ${\tilde {\cal A}}$ can be identified 
with the {\em twisted complexes} of \cite{BK}, {\em defined over ${\cal A}$}. 
The latter form a dG category, the so-called pre-triangulated category 
${\rm Pre-Tr}({\cal A})$ associated with the one-object category ${\cal A}$. 
Hence we have the identification: 
\be
p({\tilde {\cal A}})={\rm Pre-Tr}({\cal A})~~.
\ee
This is a reflection of the general principle, already mentioned in 
\cite{com3}, that 
{\em the category of pseudocomplexes of the shift-completion ${\tilde{\cal A}}$
coincides with the pre-triangulated category ${\rm Pre-Tr}({\cal A})$ of the 
uncompleted category ${\cal A}$.} From this point of view, the appearance of 
pre-triangulated categories (and, later, of triangulated categories) in 
string theory is a result of taking the shift-completion. This is the general 
formulation of the main observation made by M. Douglas in 
\cite{Douglas_Kontsevich}.

Upon taking
the zeroth BRST cohomology (which, due to the existence of shift functors, 
corresponds to working {\em on-shell}) one obtains 
the so-called {\em enhanced triangulated category} 
${\rm Tr}({\cal A})=H^0({\rm Pre-Tr}({\cal A}))$
of the original one-object category ${\cal A}$. 
As discussed in 
\cite{BK}, the category ${\rm Tr}({\cal A})$ behaves in a certain sense 
as a `derived category' of ${\cal A}$, thereby 
{\em providing an A-model 
analogue of the derived category picture familiar from studies of the 
B-model} \cite{Kontsevich, Douglas_Kontsevich, Aspinwall}. 
This construction can be related to a degeneration of the large radius limit 
of the `derived category' of 
Fukaya's category \cite{Kontsevich, Fukaya, Fukaya2, Kontsevich_recent}
(see Section 9).

\subsection{Generalized complexes and the quasiunitary cover}

We saw above that the traditional approach to A-type brane dynamics
(which largely consists of applying the results of \cite{Witten_CS} ) 
must be extended in a rather nontrivial manner. 
We hope to have convinced the reader that, 
when confronted with the problem of analyzing the structure 
defined by the most general solutions of (\ref{eq}), the tools of 
category theory become not only directly relevant, but also unavoidable, 
at least as a first approach to organizing the resulting complexity. 

It may then come as a surprise that the most general solutions of (\ref{eq}) 
{\em do not, in fact, suffice}. Indeed, it was 
shown in \cite{com1, com3} that the basic 
physical constraint of {\em unitarity} requires the consideration of even more 
general objects (the so-called {\em generalized complexes}) over ${\cal A}$, 
which can be described as `pseudocomplexes with repetition'.
In a generalized complex, one allows for a sequence $a_{n_j}$ whose 
(non-necessarily distinct) terms belong to our D-brane family $\{a_n\}$, 
and one asks for solutions of the obvious generalization of (\ref{eq}). 
More precisely, a generalized complex over ${\tilde {\cal A}}$ 
can be described as a sequence $(a_{n_j})$ together with a 
family of morphisms $\phi_{n_i,n_j}\in Hom^1(a_{n_i}, a_{n_j})$
subject to the conditions:
\be
Q_{m_i,m_k}\phi_{m_i,m_k}+\sum_{j}{\phi_{m_j,m_k}\phi_{m_i,m_j}}=0~~.
\ee
The generalization away from pseudocomplexes is due to the fact that we 
allow for repetitions $m_i=m_j$. 

It was shown in \cite{com1} that generalized complexes automatically 
lead to a string field theory, the so-called {\em quasiunitary cover}
$c({\tilde {\cal A}})$ of 
the theory based on ${\tilde {\cal A}}$. The quasiunitary cover 
satisfies a minimal 
form of the physical constraint of unitarity. Namely, any condensation 
process in the string field theory based on $c({\tilde {\cal A}})$ 
produces a D-brane which can be identified with 
an object of $c({\tilde {\cal A}})$.
Mathematically, this gives an extension of the 
Bondal-Kapranov theory, which is forced upon us due to very basic 
physical considerations. It is {\em this} theory which gives the 
ultimate (i.e. `physically closed') extension of the naive one-object 
theory described by ${\cal A}$. 

\section{Relation with the large radius limit of Fukaya's category} 
In the last section of this paper, I wish to give a short outline of the 
how the theory considered here may relate to the category constructed in 
\cite{Fukaya, Fukaya2}. As I will discuss in more detail somewhere else, 
Fukaya's category is related to a `quantized' version of a category of 
{\em intersecting} A-type branes, where the quantum effects arise from 
disk instanton corrections to the semiclassical structure. 

Since disk instanton
effects are suppressed in the large radius limit, we can in first 
approximation 
consider the case when the Kahler class of our Calabi-Yau manifold 
belongs to the deep interior of the Kahler cone. In this case, it can 
be shown that the $A_\infty$ 
`category' considered in \cite{Fukaya} reduces to a 
$\Z_2$-projection of a {\em substructure} of an 
associative differential graded category which satisfies the axioms 
of \cite{com1}. This substructure, which is not a category in 
the standard mathematical sense, only describes a {\em subsector} 
of the (large radius) open string field theory of the topological A-type string. 
More precisely, the construction proposed in \cite{Fukaya, Fukaya2} 
corresponds to a `category' whose objects are A-type branes wrapping 
{\em distinct} and {\em transversely intersecting} Lagrangian cycles, 
and whose morphisms are given by boundary condition {\em changing} states 
of the topological A-type string. This construction does 
not take into account the possibility of having different A-type branes 
wrapping the same Lagrangian cycle (which is precisely the subject of 
interest 
in the present paper), nor does it consider the boundary sectors of 
strings starting and ending on the same D-brane. Moreover, the construction 
of Fukaya's categiory 
is currently largely 
performed in a $\Z_2$-graded approach, and should be extended 
by inclusion of $\Z$-gradings.

If we use the notation 
$supp(a)$ to describe the support of a topological A-type brane, 
i.e. its underlying Lagrangian cycle, then the large radius limit 
of Fukaya's category consists 
of a collection of objects having the property that 
$supp(a)$ and $supp(b)$ are transversely intersecting Lagrangian cycles 
for any pair of {\em distinct} D-branes $a$ and $b$. Moreover, the original
proposal of \cite{Fukaya} only considers morphism spaces of the form 
$Hom(a,b)$ for $a\neq b$, i.e. no endomorphism spaces $End(a):=Hom(a,a)$ 
are allowed. Due to this reason, the resulting large radius structure 
does 
not correspond to 
a category in the classical mathematical sense, in spite of the fact that 
all off-shell nonassociativity
can be eliminated in the large radius limit. 
For example, this structure does not 
contain units $1_a\in Hom(a,a)$, since morphism spaces of the type 
$Hom(a,a)$ are not directly described. This leads to somewhat complicated 
constructions \cite{Fukaya2, Kontsevich_recent} which attempt to repair such 
problems by making use of transversality arguments
\footnote{The main idea of this approach is to view the Fukaya category as 
a description of a certain form of intersection theory for Lagrangian cycles.
In this case, the missing endomorphisms can be recovered by considering 
Lagrangian {\em self}-intersections, which can be defined in traditional 
topological manner as intersections between a Lagrangian cycle and a small 
displacement of itself (in this symplectic theory, such a displacement 
corresponds to an isotopy transform of the cycle). While one expects
such a procedure to be related to the physical approach of 
including diagonal boundary sectors from the very beggining (or at least to a 
subsector of the physical construction), a 
reasonably complete proof that this is indeed the case has not yet been given.
I thank Prof. K. Fukaya for some clarifications on the current status 
of his work on this issue.} . 

It is now clear that the theories described in this paper belong 
precisely to the missing sector of the original construction of 
\cite{Fukaya, Fukaya2}. Indeed, we studied exactly the case when 
one has distinct topological D-branes which wrap the same (special) Lagrangian
cycle. One can view part of this as a certain degeneration of the 
large radius version of Fukaya's `category' in which various Lagrangian cycles 
are deformed until they coincide. 
More precisely, it seems likely that only the 
sector of unidirectional off-diagonal 
deformations discussed in  Subsection 8.2.
can be recovered in this manner from Fukaya's category. 
However, the issue is clouded by that fact that, in a 
string field theory such as those discussed in \cite{com1, com3}, 
physics is invariant with respect to so-called {\em quasiequivalences},
and for this (as well and other) reasons the issue is currently unsettled.
A perhaps more natural point of  view is to follow 
the physics by including such objects in the very definition of 
the relevant category, as required by the structure 
of open string field theory.

\section{Conclusions and directions for further research}

We studied (the large radius limit of) a sector of 
the string field theory of the open A-model, by 
considering a system of distinct topological D-branes which wrap the 
same special Lagrangian cycle. We extracted the relevant string field 
action from first physical principles, and identified it with 
a $\Z$-graded version of super-Chern-Simons field theory, thereby 
relating graded A-brane dynamics with the mathematical theory 
of $\Z$-graded superconnections developed in \cite{Bismut_Lott}. 

Upon using the resulting string field action, we gave a preliminary 
discussion of the associated moduli space of vacua, and we sketched 
its relation it with 
the theory of {\em extended} deformations \cite{Kontsevich_Felder, Manin} 
of flat connections on the cycle. 
Moreover, we studied the effect of vacuum deformations form the 
perspective of \cite{com1, com3}, which relates them to 
condensation of boundary and boundary condition changing 
operators and to formation of D-brane composites. 
This gives an explicit realization of 
the general discussion of those papers and shows the existence 
of a large class of `exotic' A-type branes.
We also made a few observations 
about the connection of this physically motivated construction with the 
theory of enhanced triangulated categories developed in \cite{BK}. 

Our analysis should be viewed as a description of a small sector of the 
A-model counterpart of the `derived category of D-branes' whose B-model 
incarnation leads to the derived category of coherent sheaves. 
It is a basic consequence of mirror symmetry for open strings that the 
processes of D-brane composite formation which are responsible for 
generating the derived category of coherent sheaves in the B-model should 
have an A-model counterpart. Just as $D^bCoh$ can be viewed 
as the product of off-shell B-type open string dynamics 
\cite{Douglas_Kontsevich, Aspinwall, com3, us, Diaconescu}, 
A-model composite formation processes lead to an 
enlargement of the standard category of A-type topological D-branes. 
This enlargement, which after quantization ($=$ inclusion of 
disk instanton effects) can be viewed as a sort of `derived category' of a 
completed\footnote{Completed by inclusion of endomorphisms.} version of 
Fukaya's category, generates the A-model counterpart of $D^bCoh$
\footnote{Such an enlargement of Fukaya's category was already proposed 
in \cite{Kontsevich}, though the proposal of that paper does not 
consider diagonal boundary sectors.}. 

The generalized topological D-brane composites 
constructed in this paper correspond to backgrounds 
belonging to the extended moduli space of the topological A-string. 
As such, they should be relevant for a better formulation of 
homological mirror symmetry, a subject which forms the deeper motivation 
of our study. It is important, however, to approach the much harder question 
whether such objects play a role in the physical, untwisted model. 
This issue could in principle be addressed through a careful study of 
deformations for the string field theory of compactified superstrings 
in the presence of graded A-type branes.

This problem is somewhat difficult to formulate precisely,
due to the fact that current understanding of the {\em super}string 
field theory of Calabi-Yau compactifications is rather incomplete. 
Since the topological A-model only captures the chiral primary sector,  
it is in principle possible that some of the extended deformations leading 
to our exotic D-branes are `lifted' in the physical theory, due to 
the dynamics of higher string modes. It is likely, however, that at least the 
covariantly-constant complexes of Subsection 8.2. survive in the untwisted 
model, since one can adapt effective action arguments such as those of 
\cite{Oz_triples} to argue for their appearance on non-topological grounds.
In fact, these objects are the analogues of the type B D-brane complexes of 
\cite{Douglas_Kontsevich, Aspinwall}, and their relevance for 
Calabi-Yau {\em super}string dynamics should be similar
to the importance of the latter.
Whether the more exotic composites resulting from condensation of 
higher rank forms also play a role in the non-topological theory is currently 
an open problem, which deserves careful study.

\

\acknowledgments{
I wish to thank Prof. Sorin Popescu for collaboration in a related 
project and
Profs. D. Sullivan and K. Fukaya for useful conversations.
I am endebted to 
Prof. M.~Rocek for constant support and interest in my work. 
The author is supported by the Research Foundation under NSF grant 
PHY-9722101. }

\


\begin{thebibliography}{100}
\bibitem{Kontsevich}{M.~Kontsevich,{\em Homological algebra of mirror 
symmetry},  Proceedings of the International 
Congress of Mathematicians,
(Zurich, 1994), 120--139, Birkhauser, alg-geom/9411018.}
\bibitem{Seidel}{P.~Seidel, {\em Graded Lagrangian submanifolds}, 
Bull. Soc. Math. France 128 (2000), 103-149, math.SG/9903049.}
\bibitem{Douglas_Kontsevich}{M.~Douglas,
{\em D-branes, Categories and N=1 Supersymmetry}, hep-th/0011017.}
\bibitem{pi_stab}{ Michael R. Douglas, Bartomeu Fiol, Christian Romelsberger, 
{\em Stability and BPS branes}, hep-th/0002037.}
\bibitem{Aspinwall}{Paul S. Aspinwall, Albion Lawrence, 
{\em Derived Categories and Zero-Brane Stability}, 
hep-th/0104147.}
\bibitem{Zaslow_hms}{Alexander Polishchuk, Eric Zaslow, 
{\em Categorical Mirror Symmetry: The Elliptic Curve}, 
Adv.Theor.Math.Phys. {\bf 2} (1998) 443-470, math.AG/9801119.}
\bibitem{Oz_triples}{Yaron Oz, Tony Pantev, Daniel Waldram,
{\em Brane-Antibrane Systems on Calabi-Yau Spaces}, 
hep-th/0009112.}
\bibitem{Oz_superconn}{
Mohsen Alishahiha, Harald Ita, Yaron Oz, 
{\em On Superconnections and the Tachyon Effective Action}, 
hep-th/0012222.}
\bibitem{Ooguri}{ Hirosi Ooguri, Yaron Oz, Zheng Yin, {\em 
D-branes on Calabi-Yau spaces and their mirrors}, 
Nucl.Phys. {\bf B477} (1996) 407-430.}
\bibitem{top}{C.~I.~Lazaroiu, 
{\em On the structure of open-closed topological field 
theory in two dimensions}, hep-th/0010269, to be published in Nucl. Phys. B}
\bibitem{boundary}{C.~I.~Lazaroiu, 
{\em  Instanton amplitudes in open-closed topological string theory}, 
 hep-th/0011257.}
\bibitem{com1}{
C.~I.~Lazaroiu, {\em Generalized complexes and string field theory}, 
hep-th/0102122.}
\bibitem{com3}{C.~I.~Lazaroiu, {\em 
Unitarity, D-brane dynamics and D-brane categories}, 
hep-th/0102183. }
\bibitem{us}{C. I. Lazaroiu and  S. Popescu, {\em to appear}.}
\bibitem{nlsm}{C. I. Lazaroiu, in preparation}
\bibitem{Diaconescu}{D.E. Diaconescu
{\em Enhanced D-Brane Categories from String Field Theory}, hep-th/0104200.}
\bibitem{BK}{A.~Bondal, M.~M.~Kapranov, 
{\em Enhanced triangulated categories}, Mat. Sb. {\bf 181} (1990), No.5, 669, 
English translation in Math. USSR Sbornik Vol {\bf 70} (1991), No. 1 , 93. }
\bibitem{Quillen}{D. Quillen, 
{\em Superconnections and the Chern character}, 
Topology, {\bf 24}, No.1.(1085), 89-95.}
\bibitem{Bismut_Lott}{J.~M.~Bismut  and J.~Lott, 
{\em Flat vector bundles, direct images and higher analystic torsion}, 
J. Amer. Math Soc {\bf 8} (1992) 291.}
\bibitem{Vafa_cs}{C. Vafa, 
{\em Brane/anti-Brane Systems and $U(N|M)$ Supergroup},  hep-th/0101218.}
\bibitem{Kontsevich_Felder}{M.~Kontsevich, 
{\em Deformation quantization of Poisson Manifolds}, I, 
mat/9709010.}
\bibitem{Manin}{Yu. I. Manin
{\em Three constructions of Frobenius manifolds: a comparative study},
Asian J. Math. {\bf 3} (1999), no. 1, 179--220, math.QA/9801006. }
\bibitem{Barannikov_formality}{Sergey Barannikov, 
{\em  Generalized periods and mirror symmetry in dimensions $n>3$}, 
 math.AG/9903124.}
\bibitem{AS1}{Scott Axelrod, I. M. Singer,
{\em Chern-Simons perturbation theory}, 
Proceedings of the XXth International Conference on 
Differential Geometric Methods in Theoretical Physics,
Vol. 1, 2 (New York, 1991), 3--45, World Sci. Publishing, River Edge, 
NJ, 1992, hep-th/9110056.}
\bibitem{Cardy}{J.~L.~Cardy, 
{\em Conformal invariance in critical systems with boundaries}, 
Bonn 1986, Proceedings, Infinite Lie Algebras and Conformal Invariance 
In Condensed Matter and Particle Physics*,
81-92. ; {\em Boundary conditions, fusion rules and the Verlinde formula}, 
Nucl.Phys. {\bf B324} (1989) 581; {\em Boundary conditions in conformal field 
theory}, in {\em 
Integrable systems in quantum field theory and statistical mechanics}, 
eds. M. Jimbo et al, 127--148; 
J.~L.~Cardy, D.~C.~Lewellen, 
{\em Bulk and boundary operators in conformal field theory}, 
 Phys.Lett. {\bf B259} (1991), 274--278.}
\bibitem{Witten_CS}{
E.~Witten,{\em Chern-Simons gauge theory as a string theory}, 
The Floer memorial volume, 637--678, Progr. Math., 133, Birkhauser, Basel,
1995, hep-th/9207094.}
\bibitem{Witten_NLSM}{
E.~Witten, {\em Topological sigma models}, Commun. Math. Phys. 
{\bf 118} (1988),411.}
\bibitem{Witten_mirror}{E.~Witten, 
{\em Mirror manifolds and topological field theory}, 
Essays on mirror manifolds, 120--158, Internat. Press, 
Hong Kong, 1992, hep-th/9112056.}
\bibitem{Witten_SFT}{E.~Witten, {\em Noncommutative geometry and string 
field theory}, Nucl. Phys, {\bf B268} (1986) 253.}
\bibitem{Thorn}{C.~B.~Thorn, {\em String field theory},~Phys.~Rept. {\bf175}(1989)1.}
\bibitem{Zwiebach_closed}{B.~Zwiebach,
{\em Closed string field theory: Quantum action and the B-V master equation},
Nucl.~Phys. {\bf B 390}(1993) 33, hep-th/9206084.}
\bibitem{Zwiebach_open}{
B.~Zwiebach, {\em Oriented open-closed string theory revisited}, 
Annals. Phys. {\bf 267} (1988), 193, hep-th/9705241.}
\bibitem{Gaberdiel}{
M.~Gaberdiel, B.~Zwiebach, {\em 
Tensor constructions of open string theories I:Foundations
}, Nucl. Phys {\bf B505} (1997), 569, hep-th/9705038.}
\bibitem{boundary_states}{ Andreas Recknagel, Volker Schomerus,
{\em Moduli Spaces of D-branes in CFT-backgrounds},
hep-th/9903139,{\em Boundary Deformation Theory and Moduli Spaces of D-Branes},
Nucl.Phys. B545 (1999) 233-282hep-th/9811237,{\em D-branes in Gepner models},
Nucl.Phys. B531 (1998) 185-225,hep-th/9712186,
N.~Ishibashi, {\em The boundary and crosscap states
 in conformal field theories}, Mod.~Phys.~Lett. {\bf A4} (1989) 251; 
 N.~Ishibashi, T.~Onogi, {\em Conformal field theories on surfaces
 with boundaries and crosscaps}, Mod.~Phys.~Lett. {\bf A4} (1989) 161;}
\bibitem{Fukaya}{K.~Fukaya, 
{\em  Morse homotopy,  $A^\infty$-category and  Floer  homologies}, in
{\em  Proceedings of  the  GARC Workshop  on  Geometry and  Topology},
ed. by H.~J.~Kim, Seoul  national University (1994), 1-102; {\em Floer
homology, $A^\infty$-categories and topological field theory}, in {\em
Geometry and Physics}, Lecture  notes in pure and applied mathematics,
{\bf 184},  pp 9-32, Dekker, New  York, 1997; {\em  Floer homology and
Mirror       symmetry,      I},       preprint       available      at
$http://www.kusm.kyoto-u.ac.jp/~\tilde{}~fukaya/fukaya.html.$}
\bibitem{Fukaya2}{K. Fukaya, Y.-G. Oh, H.~Ohta, K.~Ono, 
{\em Lagrangian intersection Floer theory - anomaly and obstructon}, 
  preprint       available      at
$http://www.kusm.kyoto-u.ac.jp/~\tilde{}~fukaya/fukaya.html.$}
\bibitem{Kontsevich_recent}{M.~Kontsevich, Y.~Soibelman, 
{\em Homological mirror symmetry and torus fibrations}, math.SG/0011041.}
\end{thebibliography}
\end{document}